\documentclass[aps,prx,twocolumn,showpacs]{revtex4}
\usepackage{graphicx}
\usepackage{epstopdf} 
\usepackage{amsmath}
\usepackage{amsfonts}
\usepackage{lipsum}
\usepackage{makecell}

\newcommand{\beq}{\begin{equation}}
\newcommand{\enq}{\end{equation}}
\newcommand{\bea}{\begin{eqnarray}} 
\newcommand{\ena}{\end{eqnarray}}

\begin{document}
 
\title{Dynamics of non-equilibrium steady state quantum phase transitions}
\author{Patrik Hedvall}
\author{Jonas Larson}
\affiliation{Department of Physics, Stockholm University, AlbaNova
  University Center, Se-106 91 Stockholm, Sweden}

\date{\today}

\begin{abstract}
In this paper we address the question how the Kibble-Zurek mechanism, which describes the formation of topological defects in quantum systems subjected to a quench across a critical point, is generalized to the same scenario but for driven-dissipative quantum critical systems. In these out of equilibrium systems, the critical behavior is manifested in the steady states rather than in the ground states as in quantum critical models of closed systems. To give the generalization we need to establish what is meant by adiabaticity in open quantum systems. Another most crucial concept that we clarify is the characterization of non-adiabatic excitations. It is clear what these are for a Hamiltonian systems, but question is more subtle for driven-dissipative systems. In particular, the important observation is that the instantaneous steady states serve as reference states. From these the excitations can then be extracted by comparing the distance from the evolved state to the instantaneous steady state. With these issues resolved we demonstrate the applicability of the generalized Kibble-Zurek mechanism to an open Landau-Zener problem and show universal Kibble-Zurek scaling for an open transverse Ising model. Thus, our results support any assumption that non-equilibrium quantum critical behavior can be understood from universal features.
\end{abstract}

\maketitle 

\section{Introduction}\label{sec:intro}
The theory for describing systems at equilibrium, and especially what drives transitions between different phases, is well developed and the picture rather complete, whether it is for finite temperature (thermal phase transition (PT))~\cite{goldenfeld1992lectures}, or at zero temperature (quantum phase transition (QPT))~\cite{sachdev2007quantum}. A great leap in our understanding was taken with Landau's mean-field theory of phase transition, which in particular tells that a continuous (i.e. second order) PT can be ascribed a spontaneous breaking of some symmetry~\cite{landau1937theory}. The type of symmetry being broken also predicts what kind of excitations one can predict in the symmetry broken phase, so called Higgs or Goldstone modes representing respectively massive or massless excitations~\cite{altland2010condensed}. The non-analytic behavior at a quantum critical point implies that the spectrum is necessarily gapless at this point, and more importantly the physics in the vicinity of the critical point, for example how the gap closes, is universal. This means that microscopic details become irrelevant and the physics can be described by a set of critical exponents that only depend on global symmetries and dimensions~\cite{goldenfeld1992lectures}. The fact that the gap closes means that the typical time-scale (inverse of the gap) diverges, what is called critical slowing down. An interesting and important result of critical slowing down is that if we drive the system through a critical point, by externally varying some system parameter, no matter how slow this quench is the system will always get excited due to non-adiabatic contributions. Kibble and independently Zurek suggested that the density of these generated defects also behaves universal -- it can be estimated from some critical exponents~\cite{kibble1976topology,zurek1985cosmological}. While originally for thermal PT's, the idea of Kibble and Zurek can be applied to QPT's as well~\cite{zurek2005dynamics}. This idea is simple, the evolution of the quench can be divided into being either adiabatic (typically when the gap is large) or diabatic (in the vicinity of the critical point where the gap closes). This has been termed the Kibble-Zurek mechanism (KZM), and its applicability has been demonstrated in numerous experiments by now~\cite{bowick1994cosmological,ruutu1996vortex,ducci1999order,monaco2002zurek,maniv2003observation,weiler2008spontaneous,bakr2010probing,chen2011quantum,braun2015emergence,ulm2013observation,pyka2013topological,mielenz2013trapping}.  

Lately a new type of critical behavior in non-equilibrium systems has gained much attention thanks to great experimental advances especially in the AMO (atomic, molecular and optical) community. These are driven-dissipative systems, in which, in principle, both the drive and the sort of dissipation can be tailored~\cite{diehl2008quantum,verstraete2009quantum,eisert2010noise,hoening2012critical}. The system relaxes to some steady state that will not be an equilibrium one, i.e. a non-equilibrium steady state or short NESS. Criticality in these models are correspondingly defined via non-analyticities in the system's steady state $\hat\rho_\mathrm{ss}$ upon varying some model parameter, similar to how the ground state $|\psi_0\rangle$ shows non-analytic behavior at the critical point for a QPT. For AMO experiments, the evolution of the system in contact with an environment can often be accurately described by a Markovian Lindblad master equation~\cite{gardiner2015quantum,breuer2002theory}   
\begin{equation}\label{lindblad}
\begin{array}{lll}
\partial_t\hat\rho(t) & = & \displaystyle{\hat{\mathcal L}(\hat\rho(t))\equiv i\left[\hat\rho(t),\hat H\right]}\\ \\ & & \displaystyle{+\sum_i\kappa_i\left(2\hat L_i\hat\rho(t)\hat L_i^\dagger-\hat L_i^\dagger\hat L_i\hat\rho(t)-\hat\rho(t)\hat L_i^\dagger\hat L_i\right).}
\end{array}
\end{equation}
The influence of the environment is included in the last term containing a sum over possible Lindblad jump operators $\hat L_i$, and where the $\kappa_i$'s are the corresponding decay rates. When tailoring the system-environment couplings these are constructed with the purpose of reaching a desired steady state $\hat\rho_\mathrm{ss}$ of Eq.~(\ref{lindblad}). Typically the jump operators are local but in principle need not be so~\cite{schneider2002entanglement,hannukainen2017dissipation}. For $\hat\rho_\mathrm{ss}$ to be non-analytic in the thermodynamic limit, the gap in the spectrum, now of the Liouvillian $\hat{\mathcal L}$, must close at the critical point~\cite{kessler2012dissipative}. Since the time-evolution governed by $\hat{\mathcal L}$ is in general not unitary, the spectrum is normally complex with the real parts representing relaxation to the NESS's~\cite{albert2014symmetries,albert2016geometry}. 

It is clearly so that our understanding of NESS critical behavior is far less developed than what it is for thermal or quantum PT's. It has been shown, for example, that new universality classes are possible in these systems~\cite{sieberer2013dynamical,marino2016driven,zamora2017tuning}, as is the presence of continuous PT's lacking any spontaneous symmetry breaking~\cite{hannukainen2017dissipation}. Hence, qualitative difference between equilibrium and non-equilibrium QPT's may indeed exist. A very famous example from classical physics is the `flocking transition' which describes build-up of long-range order in for example flocks of birds~\cite{vicsek1995novel}. At equilibrium, long-range order is prohibited by the Mermin-Wagner theorem~\cite{auerbach2012interacting}, but the theorem does not apply for non-equilibrium situations and thereby the possibility for the birds to order. 

In the present paper we analyze how NESS critical systems, described by some Lindblad master equations, respond to slow quenches across a critical point. As pointed out above, for equilibrium systems we know that under rather general circumstances the KZM accurately explains such scenarios. Thus, the corresponding question is whether some KZM can be applied also for these models? We will see that there is indeed a generalization of the traditional KZM to open systems, but much care needs to be taken into account to settle this correspondence. For example, how the concept of adiabaticity is translated to open quantum systems, and more importantly how to quantify the amount of non-adiabatic excitations generated during the quench. It is clear that since the criticality manifests in the system's steady state, this state should be the reference when determining the amount of excitations. We will argue that the natural measure for the amount of excitations is the trace distance for density operators.   

It is in general harder to solve dynamical than statical or equilibrium problems, and in particular it is more difficult to simulate the evolution governed by a Lindblad master equation than that deriving from a Hamiltonian. This is easily understood due to the simple observation that one needs many more values to describe a general state $\hat\rho(t)$ than a pure state $|\psi(t)\rangle$. Indeed, for a Hilbert space dimension $D$, the number of eigenstates of $\hat{\mathcal L}$ is $D^2$ to be compared to $D$ eigenstates of $\hat H$. A zero eigenvalue eigenstate $\hat\rho_\mathrm{ss}$ is obviously a steady state of the Liouvillian. Adiabatic evolution should imply that a system initialized in some steady state $\hat\rho_\mathrm{ss}(t_i)$ at time $t_i$ remains in the instantaneous steady state $\hat\rho_\mathrm{ss}(t)$ throughout the duration of the quench. In the past, several works have studied how an environment affects the KZ-type quench through a quantum critical point~\cite{fubini2007robustness,cincio2009dynamics,dutta2016anti,dziarmaga2006dynamics}. This is a conceptually different question than the one we address: In our setting we cannot in general characterize the system from some underlying Hamiltonian -- it is truly the system plus environment that determines the properties of the state $\hat\rho(t)$. In the Refs.~\cite{fubini2007robustness,cincio2009dynamics,dutta2016anti,dziarmaga2006dynamics} the environment serves mainly as an additional source of fluctuations, and the general finding is that the amount of excitations created when quenched through the (Hamiltonian) critical point is increased. It is, in particular, assumed that the presence of an environment will not qualitatively alter the critical behavior of the Hamiltonian. Moreover, excitations are measured with respect to the Hamiltonian ground state, and not from the steady state as in the present work. Another crucial observation is that when the ground state is the reference state, as in the mentioned references, the amount of excitations will not, in a strict sense, only depend on the quench rate and critical exponents but also on the initial and final times, $t_i$ and $t_f$. This is irrespective of whether the quench end close or far from a critical point, since even far from the critical point, where the evolution is presumably adiabatic, there is an environment induced relaxation of the system towards some steady state. In our analysis of quenches through NESS critical points, the results do not depend on $t_i$ and $t_f$, and thereby only rely on system parameters and properties. In the respect, the approach we take is more appropriate when the goal is to explore universal properties of the critical behavior.

The paper is organized as follows. In the following section we summarize some earlier results such that they can be put in context with ours, and we also classify different scenarios that can emerge in NESS critical systems. More precisely; in Sec.~\ref{ssec:KZM} we reproduce the arguments (so called `adiabatic-impulse approximation') for the KZM when applied to quantum critical points of closed systems, then in Sec.~\ref{ssec2B} we continue with defining NESS criticality and introduce different classes arising from the competing terms in the Lindblad master equation, and in Sec.~\ref{ssec2C} we review earlier works related to our results. Section~\ref{sec3} presents the general results. We start in Sec.~\ref{ssec3A} by introducing the Bloch equations of Lindblad master equations and discuss their structure in rather general terms. This allows us to explain what is meant with adiabaticity for open quantum systems in Sec.~\ref{adsubsec}. To use the trace distance as a measure of excitations is argued for in Sec.~\ref{adsubsec} by demonstrating that it correctly predicts the amount of excitations in the limiting situation of a closed quantum system. In Sec.~\ref{ssec3d} we generalize the adiabatic-impulse approximation to our Bloch equations and thereby derive the KZM for open quantum systems. The proceeding Sec.~\ref{sec4} is devoted to two examples, the open Landau-Zener model in Sec.~\ref{ssec4A} and the dephasing transverse Ising model in Sec.~\ref{ssec4B}. The two examples verify the applicability of our general results from Sec.~\ref{sec3}. We conclude in Sec.~\ref{sec5} with a summary and possible future continuations.


\section{Preliminaries and earlier results}
\subsection{Dynamics of quantum phase transitions}\label{ssec:KZM}
In this Subsection we reproduce the idea of the adiabatic-impulse (AI) approximation, and how the KZM is applied to QPT's~\cite{zurek2005dynamics}. A QPT occurs at zero temperature and result from quantum fluctuations~\cite{sachdev2007quantum}, contrary to classical PT's which are driven by thermal fluctuations. Typically for a system showing critical behavior, its Hamiltonian can be decomposed as
\begin{equation}
\hat H=\hat H_0+\lambda\hat H_1,
\end{equation}
with $\left[\hat H_0,\hat H_1\right]\neq0$ and $\lambda$ is a coupling parameter determining the relative strength between the two terms. For $\lambda=0$, the ground state is governed by $\hat H_0$, while in the opposite limit, $\lambda\rightarrow\infty$, it is dictated by $\hat H_1$. The characteristics of the ground state $|\psi_0(\lambda)\rangle$ may be very distinct in the two limits, and in particular in the thermodynamic limit it may become non-analytic for some finite critical coupling $\lambda_c$. When the transition is continuous this marks the critical point. The non-analytic behavior necessarily implies that the spectrum of $\hat H$ is gapless at $\lambda_c$, i.e. the energy gap from the ground state  $\Delta_H\rightarrow0$. In particular, for $\lambda<\lambda_c$ the ground state is non-degenerate and the spectrum gapped (normal phase), while the ground state is degenerate  for $\lambda>\lambda_c$ (symmetry broken phase) and the spectrum may either be gapped or not. At the critical point the ground state cannot be said to be ruled by either $\hat H_0$ or $\hat H_1$, and furthermore we cannot assign any finite characteristic length scale to the state. In the vicinity of the critical point, the length scale diverges algebraically~\cite{goldenfeld1992lectures}
\begin{equation}
\xi\propto|\lambda-\lambda_c|^{-\nu},
\end{equation}
for some universal critical exponent $\nu$ that does not depend on microscopic details but only on global symmetries and dimensions. The inverse of the gap, $\Delta_H^{-1}$, sets an intrinsic time-scale for  how fast the system responds to small parameter changes. Since the behavior of the gap closing $\Delta_H\rightarrow0$ is universal we must have that this time-scale also diverges with some dynamical exponent $z$ defined as  
\begin{equation}\label{dyncrit}
\tau_H\propto|\lambda-\lambda_c|^{-\nu z}.
\end{equation}

\begin{figure}
\includegraphics[width=8cm]{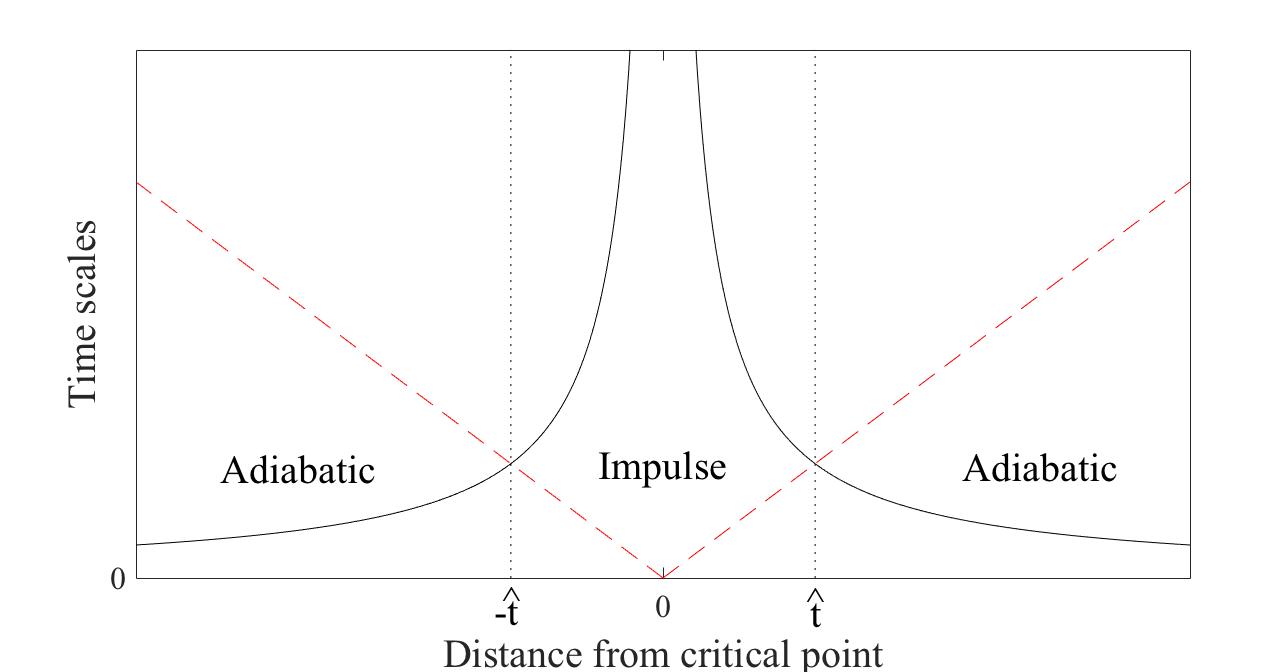} 
\caption{The idea behind the AI-approximation which forms the basis for the KZM applied to QPT's. Far from the critical point, the energy gap is large such that the reaction time $\tau_H$ (marked by solid black lines) is small, i.e. the system responds fast to any external changes. In this adiabatic regime the inverse transition rate $\varepsilon/\dot\varepsilon$ (dashed red lines), giving the time-scale for the external driving, is much larger than the reaction time. As the critical point is approached, the reaction time becomes longer and at the freeze-out time $-\hat t$ the system cannot stay adiabatic and it enters the impulse regime. Here the evolution is assumed frozen, i.e. diabatic. After traversing the critical point, the system can reenter into an adiabatic regime at the second freeze-out time $\hat t$. 
}
\label{fig1}
\end{figure}
  
We assume now that the system is quenched through the critical point such that~\cite{zurek2005dynamics}
\begin{equation}\label{quench}
\varepsilon(t)\equiv\lambda(t)-\lambda_c=-\frac{t}{\tau_Q},
\end{equation}
where $\tau_Q$ is the `quench rate', i.e. it sets the rate of change of the Hamiltonian. At the time $t=0$ we are exactly at the critical point, and we assume that for negative times the instantaneous ground state is the gapped normal phase. Even though the energy gap closes at the critical point, we may assume that sufficiently far from the critical point the evolution is adiabatic. As we move closer to the critical point, non-adiabatic excitations cannot be overlooked. While the Hamiltonian changes smoothly, in the AI-approximation one assumes that there is a certain time $-\hat t$ for which the system swaps from following the quench adiabatically to diabatically. Thus, after this {\it freeze-out time} the population transfer is hindered -- the evolution is frozen. The design is thereby such that the quench can be split into an adiabatic regime followed by an {\it impulse} regime -- the AI-approximation. Within the AI-approximation, the problem becomes essentially one of determining $\hat t$. The breakdown of adiabaticity should occur in the vicinity of the point when the rate of change $|\dot\varepsilon(t)/\varepsilon(t)|$ equals the response time $\tau_H=\Delta_H^{-1}$. Using (\ref{dyncrit}) and the explicit form of the quench (\ref{quench}), one finds
\begin{equation}
\hat t\sim\tau_Q^{\frac{\nu z}{1+\nu z}}.
\end{equation}
The validity of the AI-approximation hinges on that the window where the evolution is neither adiabatic nor diabatic is narrow. Typically, the quench rate should not be too small for good predictions from the KZM, since then this window is not narrow any more.

Up to the instant $-\hat t$ we have assumed adiabatic following, and thereafter frozen evolution. The quench thereby imprints a characteristic length scale
\begin{equation}\label{length}
\xi(\hat t)\sim\varepsilon(\hat t)^{-\nu}\sim\tau_Q^{\frac{\nu}{1+\nu z}}
\end{equation}
to the system state. At $t=-\hat t$ the system is still approximately in its instantaneous ground state, but this is not the ground state at later times meaning that the system gets excited as time progresses. These are the non-adiabatic excitations. Depending on the dimensionality and the symmetries, the excitations can have different character like domain walls/kinks, or vortices. Hence, local topological defects in the otherwise uniform ground state. The length scale (\ref{length}) determines the density of such defects, i.e.
\begin{equation}\label{defdens}
n_\mathrm{D}\sim\xi^{-d}\sim\tau_Q^{-\frac{d\nu}{1+\nu z}},
\end{equation}
where $d$ is the dimension.

The KZM predicts universal behavior away from equilibrium, the scaling of the defects is determined by the critical exponents $\nu$ and $z$~\cite{del2014universality}. For a continuous PT that breaks a discrete symmetry, in the symmetry broken phase the ground state is degenerate but there is a gap to higher excited states (Higgs mode)~\cite{altland2010condensed}. The adiabatic-impulse scenario is then that you cross from adiabatic to diabatic at $-\hat t$, and then back to adiabatic at $\hat t$, see Fig.~\ref{fig1}. The imprinted excitations are still the same in this scheme. We may note that nothing restricts us to linear quenches~(\ref{quench}), and neither to the case where the two freeze-out times are symmetric, i.e. $-\hat t$ and $+\hat t$, which could for example occur when the critical exponents are different above and below the critical point~\cite{liu2009large,mumford2015dicke}. Furthermore, the transition does not have to be a proper continuous PT, but can be a crossover that appears in finite systems~\cite{zurek2005dynamics,dziarmaga2005dynamics,cincio2007entropy,de2010quench}. Indeed, the KZM can successfully be applied also to avoided crossing models~\cite{damski2005simplest,damski2006adiabatic}, which ca be seen as an approximation of a QPT in a finite system~\cite{zurek2005dynamics}. 

\begin{figure}
\includegraphics[width=8cm]{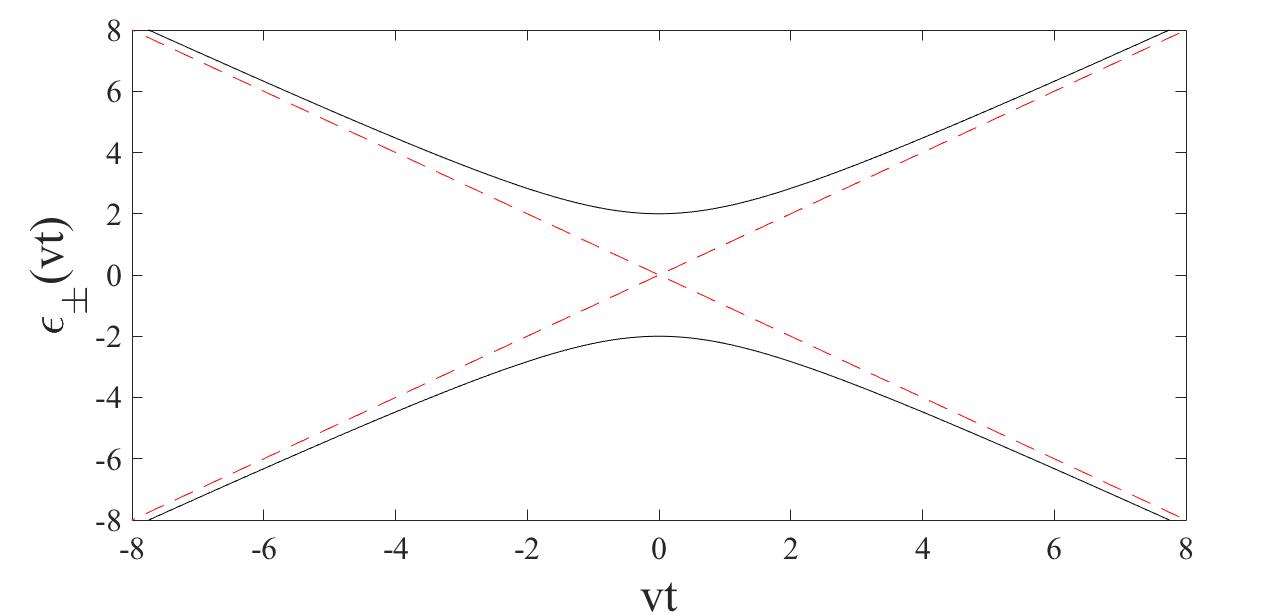} 
\caption{Adiabatic $\epsilon_\pm^{(\mathrm{ad})}(vt)=\sqrt{(vt)^2+g^2}$ (solid black lines) and diabatic $\epsilon_\pm^{(\mathrm{d})}(vt)=\pm vt$ (dashed red lines) energies of the LZ model~(\ref{lzham}) with $g=2$. The slope of the curves away from the crossing is determined by the velocity $v$, while the gap at $t=0$ is $\Delta_H(t=0)=2g$. A large coupling $g$ thus implies a large gap which favors adiabaticity, and likewise a small $v$ means a gradual change and more adiabatic evolution. 
}
\label{fig2}
\end{figure}

In Refs.~\cite{damski2005simplest,damski2006adiabatic} Damski implemented the KZM on the solvable LZ problem~\cite{zener1932non,landau1932theorie} that describes an avoided level crossing. The LZ Hamiltonian is given by ($\hbar=1$)
\begin{equation}\label{lzham}
\hat H_\mathrm{LZ}(t)=\left[
\begin{array}{cc}
vt & g\\
g & -vt
\end{array}\right],
\end{equation}
such that the quench rate $\tau_Q=v^{-1}$, and $g$ is the coupling between the two diabatic states $|0\rangle=[1\,\, 0]^T$ and $|1\rangle=[0\,\, 1]^T$. If we assume that the system starts in its ground state $|0\rangle$ for $t=-\infty$, then the probability that it ends up in the excited state at $t=+\infty$ is given by the LZ formula~\cite{zener1932non,landau1932theorie}
\begin{equation}\label{lzformula}
P_\mathrm{LZ}=e^{-\pi\frac{g^2}{v}}. 
\end{equation}
The adiabaticity parameter $\Lambda\equiv\pi g^2/v$ determines how adiabatic the process is; the larger it is the more adiabatic quench. The instantaneous eigenvalues $\epsilon_\pm^{(\mathrm{ad})}(t)=\sqrt{(vt)^2+g^2}$ of $\hat H_\mathrm{LZ}(t)$ (adiabatic energies), as well as the `bare energies' $\epsilon_\pm^{(\mathrm{d})}(t)=\pm vt$ (diabatic energies) are displayed in Fig.~\ref{fig2}. Damski found that the KZM works very well to reproduce the correct analytical result~(\ref{lzformula}) for the excitations. The agreement was particularly good for fast quenches, i.e. in the non strictly adiabatic regime~\cite{hwang2015quantum}. Naturally, there is no length scale in the LZ problem, and instead the amount of excitations $P_\mathrm{LZ}$ replaces the density of defects~(\ref{defdens}).

\subsection{NESS criticality and the Lindblad master equation}\label{ssec2B}
The conventional wisdom is that dissipation and/or decoherence decimate quantum properties~\cite{schlosshauer2007decoherence}. This is, for example, the main hindrance when it comes to quantum information processing. One possibility to circumvent this problem for quantum computing is to employ {\it decoherence-free subspaces}~\cite{lidar1998decoherence}, which uses states that are insensible for the decoherence. Relatedly, one could imagine to monitor the dissipation/decoherence channels such that the system approaches a steady state manifold where the computations take place~\cite{kraus2008preparation,diehl2008quantum,verstraete2008quantum}. Typically, these are controlled driven-dissipative systems with desirable non-equilibrium steady states $\hat\rho_\mathrm{ss}$.

Naturally, the NESS $\hat\rho_\mathrm{ss}$ may also show non-analytic behavior in some proper thermodynamic limit, just like equilibrium states can become critical~\cite{diehl2010dynamical,eisert2010noise,tomadin2010signatures,hoening2012critical,sieberer2013dynamical}. As manifestly non-equilibrium, these NESS phase transitions need not necessarily obey   laws applicable to equilibrium PT's like the Mermin-Wagner theorem~\cite{auerbach2012interacting}. There are three possible scenarios we can imagine: 

\begin{enumerate}
\item The model is critical at the closed level, i.e. the Hamiltonian supports different phases. We exclude here the possibility of new phases appearing due to the openness, see below. The questions are, how does the inclusion of dissipation/decoherence (noise) affect the types of transitions and the properties of the phases? It has been demonstrated that the presence of environmental noise can indeed change the critical exponents of the PT~\cite{nagy2011critical,baumann2011exploring}, or even lead to new universality classes~\cite{sieberer2013dynamical,marino2016driven,kardar1986dynamic,zamora2017tuning}, as well as altering the properties of the phases~\cite{joshi2013quantum,ludwig2013quantum,altman2015two}. In certain cases, the transition becomes classical and the fluctuations induced by the environment can be described as an effective temperature~\cite{mitra2006nonequilibrium,diehl2008quantum,dalla2010quantum,sieberer2013dynamical,kessler2012dissipative}, which may prohibit build-up of long-range order in lower dimensions according to the Mermin-Wagner theorem. In realistic experimental settings considering finite systems, the length scale may, however, be larger than the system size and the characteristics of the PT could in these cases still be accessible~\cite{altman2015two,dagvadorj2015nonequilibrium}.

\item Much less studied is the situation where the criticality emerges as an interplay between different dissipative channels, and not due to the Hamiltonian. Then, depending on which channel that dominates the state approaches different steady states. Verstrate {\it et al.} showed that it is possible to perform quantum computing tasks using solely dissipative evolution~\cite{verstraete2008quantum}. The target state is then a steady state and the dissipation is monitored in such a way that the desired target state is reached. 

\item The last scenario is when the criticality is only possible in the presence of both a Hamiltonian and dissipative channels~\cite{diehl2008quantum,diehl2010dynamical,eisert2010noise,hoening2012critical,hannukainen2017dissipation,carmichael2015breakdown}. Here, the emerging phases are typically resulting from either dissipation or unitary Hamiltonian evolution. The separate parts alone, Hamiltonian or dissipator, may well be trivial. It was also recently shown that the criticality in these systems can be conceptually different from equilibrium situations where a continuous PT is generally accompanied by a spontaneous symmetry breaking~\cite{hannukainen2017dissipation}.
\end{enumerate}

AMO systems have the advantage that they can be fairly versatile and offer high controllability~\cite{bloch2008many,cirac2012goals} in comparison to other material systems. The atoms or ions are manipulated with optical light, and controlled dissipation is accomplished by combination of excitations and spontaneous emission. In these optical settings, a Born-Markovian master equation is typically applicable~\cite{gardiner2015quantum}. Under such circumstances and invoking the secular (also called rotating wave) approximation, the system is described by the Lindblad master equation of Eq.~(\ref{lindblad})~\cite{breuer2002theory}. The first commutator term on the right hand side of Eq.~(\ref{lindblad}) comprises the Hamiltonian evolution, with the notification that $\hat H$ in principle also includes Lamb shifts steaming from coupling with the environment. The second term accounts for the coupling to the environment, where $\kappa_i$ are decay rates or coupling strengths, and $\hat L_i$ are the Lindblad jump operators. We have also defined the Liouvillian $\hat{\mathcal L}$ which is the total generator of the time-evolution. In what follows we will always assume that the system evolution is generated by a master equation on the form (\ref{lindblad}). 

For our purpose when studying NESS criticality, the steady state 
\begin{equation}
\hat{\mathcal L}(\hat\rho_\mathrm{ss})=0
\end{equation}
is the main objective, i.e. the kernel of the Liouvillian or the eigenstate with zero eigenvalue. It serves the same role as what the ground state does in equilibrium QPT's. The manifold $\mathcal{M}=\left\{\hat\rho\, |\,\partial_t\hat\rho=\hat{\mathcal{L}}(\hat\rho)=0\right\}$ of steady states forms a connected convex set. If the Liouvillian is time-independent, any initial state approaches a steady state in the infinite time limit~\cite{alicki286quantum,rivas2012open}. Thus, if the steady state is unique it is attractive in the meaning that any initial state will eventually path its way to it. Attractiveness implies robustness -- if for some reason the steady state is perturbed, the dynamics will bring it back. For systems with more than one steady state, attractiveness is lost as you can be brought back to another steady state. As will become clear, this observation is very important for our work. The uniqueness of steady states has been explored in the past~\cite{spohn1977algebraic,schirmer2010stabilizing,frigerio1978stationary,fagnola2001existence,fagnola2002subharmonic,baumgartner2008analysis2}. To say something general about whether the steady state will be unique or not is not trivial apart from special cases. For example, if the only operator commuting with all jump operators $\hat L_i$ is the identity then the steady state is unique~\cite{spohn1977algebraic,schirmer2010stabilizing}. It can be proven that there must exist at least one steady state~\cite{rivas2012open}. Reversely, given a pure state it is always possible to construct a Liouvillian with at least one jump operator that has this pure state as unique steady state. If the steady state is unique the evolution is ofter called relaxing. 

A general Liouvillian eigenstate has a complex eigenvalue $\mu_i$, i.e. $\hat{\mathcal L}(\hat\rho_i)=\mu_i\hat\rho_i$, such that time-evolution results in $\hat\rho_i(t)=\exp(\mu_it)\hat\rho_i$, or that we can expand a general state as
\begin{equation}\label{expand}
\hat\rho(t)=\sum_ic_ie^{\mu_it}\hat\rho_i.
\end{equation}
The eigenstates may not be physical states, i.e. positive semi-definite, and the set $\{\hat\rho_i\}$ is over-complete. Of course, Liouvillian evolution preserves positivity and norm of the state, but nevertheless the sum (\ref{expand}) may well contain unphysical states even though $\hat\rho(t)$ is physical. In fact, the eigenvalues/eigenstates typically come in complex/hermitian conjugated pairs which assures that the expansion (\ref{expand}) always exists for any physical state.

It follows that the eigenvalues must obey $\mathrm{Re}(\mu_i)\leq0$~\cite{kessler2012dissipative}. The quantity
\begin{equation}\label{lgap}
\Delta_M=\underset{i}{\mathrm{min}}\,\mathrm{Re}(-\mu_i),
\end{equation}
with the minimum taken over all eigenvalues with non-zero real parts, defines the Liouvillian gap that sets the inverse time-scale for reaching the steady state~\cite{albert2014symmetries,albert2016geometry}. The subscript is introduced to distinguish this gap from the Hamiltonian energy gap $\Delta_H$. In order for $\hat\rho_\mathrm{ss}$ to become non-analytic in the thermodynamic limit, and thereby allow for critical behavior, one must have that $\Delta_M\rightarrow0$~\cite{kessler2012dissipative}, analogously to the gap closening for equilibrium QPT's. Scaling (i.e. universality) of $\Delta_M$ is not known, and, in general, it seems that the mechanisms behind NESS QPT's can be qualitatively different from those of equilibrium QPT's~\cite{hannukainen2017dissipation}. 

In order to discuss various scenarios for different Liouvillians, and thereby try to classify them, let us assume that the sum in (\ref{lindblad}) is restricted to a single term, i.e.  we have just a single jump operator $\hat L$. This is, of course, a special case of the general situation but, as we will see, it can be a relevant case for engineered driven-dissipative systems. Even when confine the discussion to this special case, it will provide insight also for the general cases. We note also that for several jump operators $\hat L_i$, the decomposition of the Liouvillian is known to be not unique~\cite{baumgartner2008analysis}, which makes a general classification much more complicated. For single jump operators we can construct the following four classes:

\begin{description}

  \item[$\bullet$ Class I] {\it Energy dephasing}. The jump operator is hermitian, $\hat L=\hat L^\dagger$, and commutes with the Hamiltonian, $\left[\hat L,\hat H\right]=0$. Any energy eigenstate $\hat\rho_n=|E_n\rangle\langle E_n|$ will also be a steady state, and so will any mixed state $\hat\rho_{p_n}=\sum_np_n\hat\rho_n$, i.e. $\hat{\mathcal L}(\hat\rho_{p_n})=0$. Thus, the steady states are diagonal in the energy eigenbasis $\{|E_n\rangle\}$. For any initial state, evolution will cause $\hat\rho$ to become diagonal in the energy eigenbasis without dissipation of energy, but an increase of entropy~\cite{nielsen2010quantum}. The probability distribution $p_n$ determines the steady state, and, in particular, this exemplify how the manifold of steady states is convex and simply connected. 
  
   \item[$\bullet$ Class II] {\it General dephasing}. The jump operator is hermitian, $\hat L=\hat L^\dagger$, and does not commute with the Hamiltonian, $\left[\hat L,\hat H\right]\neq0$. The maximally mixed state $\hat\rho_\mathrm{ss}=\mathbb{I}/D$, with $D$ the hilbert space dimension, is a trivial steady state in this class. In many physically relevant situations, the maxiamally mixed state is also the unique steady state. With $|\varphi_n\rangle$ the eigenstates of $\hat L$, any diagonal state $\hat\rho=\sum_np_n|\varphi_n\rangle\langle\varphi_n|$ is transparent to the environment, but the Hamiltonian will drive you out of this manifold which implies relaxation to some true steady state. 
    
   \item[$\bullet$ Class III] {\it Dissipation I}. The jump operator is non-hermitian, $\hat L\neq\hat L^\dagger$, but commutes with the Hamiltonian, $\left[\hat L,\hat H\right]=0$. This class is probably the least physically relevant; we know of no non-trivial physical scenarios were it occurs. A trivial situation is that the jump operator acts in a space with only degenerate energy eigenstates, e.g. $\hat H\propto\mathbb{I}$.
         
    \item[$\bullet$ Class IV] {\it Dissipation II}. The jump operator is non-hermitian, $\hat L\neq\hat L^\dagger$, and does not commute with the Hamiltonian, $\left[\hat L,\hat H\right]\neq0$.  Non-hermitian operators typically appear in cases of spontaneous decay (e.g. $\hat L=\hat a$ for decay of photons/phonons and $\hat L=\hat\sigma^-$ for spontaneous decay of the upper level of a two-level system) or for incoherent pumping (e.g. $\hat L=\hat a^\dagger$ for the situation of a single boson mode). A dark state is a pure state $\hat\rho_{D}=|D\rangle\langle D|$ that is transparent to the dissipation/decoherence, e.g. $\hat L|D\rangle=0$~\cite{arimondo1996v}. It is clear that if further $|D\rangle$ is an eigenstate of the Hamiltonian, then $\hat\rho_{D}$ is also a steady state. The most common example being a system coupled to a zero temperature bath which cools the system down to its ground state. More generally, if the thermal bath is at some non-zero temperature, detail balance between incoherent loss and gain of particles leads to a thermal steady state provided the system is not driven. 

\end{description}

As a final remark on properties of steady states, it is rather straight forward to show~\cite{dietz2003memory} that one steady state solution falling in the classes I and III (and possibly also in the other classes in certain cases) is
\begin{equation}
\hat\rho_{\hat L}=\hat R/\mathrm{Tr}[\hat R],
\end{equation}
where $\hat R=\left(\hat L^\dagger\hat L\right)^{-1}$. 

\subsection{Earlier studies}\label{ssec2C}

\subsubsection{Dynamics across equilibrium quantum critical points}
For thermal PT's, the KZM have been well explored both numerically~\cite{laguna1997density,yates1998vortex,hindmarsh2000defect} and experimentally~\cite{bowick1994cosmological,ruutu1996vortex,ducci1999order,monaco2002zurek,maniv2003observation,weiler2008spontaneous}. It is only recently, however, that the dynamics emerging from quenches through quantum critical points has been investigated~\cite{zurek2005dynamics,dziarmaga2005dynamics,damski2005simplest,polkovnikov2005universal,schutzhold2006sweeping,cherng2006entropy,cincio2007entropy,cucchietti2007dynamics}. Laser cooling and manipulation of atoms or ions have opened up a new avenue for studying quantum many-body systems out of equilibrium~\cite{bloch2008many}. Especially the high control of system parameters make these systems good candidates for exploring critical dynamics like the KZM and its predicted scaling~\cite{dziarmaga2010dynamics,del2014universality}. 

The KZM was theoretically studied for the superfluid-to-Mott insulator transition of an ultracold atomic gas in an optical lattice~\cite{schutzhold2006sweeping,cucchietti2007dynamics}, and a KZ scaling was suggested. Quenching across this transition has also been studied in several experiments ~\cite{bakr2010probing,chen2011quantum,braun2015emergence}. In the first experiment the KZ scaling was however not analyzed. In Ref.~\cite{chen2011quantum} the scaling after quenching from the Mott to the superfluid was mapped out, but deviations from simple theory was found which was explained as a result of the inhomogeneity of the atomic cloud~\cite{dziarmaga2010dynamics,bernier2011slow,dziarmaga2014quench}. It is also known that the KZM can fail as the quench becomes too slow, where instead different scaling exponents can be predicted from adiabatic perturbation theory~\cite{hwang2015quantum}. The behavior after quenching across a PT in a spinor BEC has also been experimentally explored~\cite{sadler2006spontaneous,nicklas2015observation,anquez2016quantum}, where indeed scaling in agreement with theory was found~\cite{anquez2016quantum}. 

For trapped ions, it was suggested that the KZM could be ideally studied in the so called `zig-zag' transition in which a linear chain of trapped ions reorganize to form a zig-zag structure as the confining trapping is weakend~\cite{del2010structural,de2010spontaneous}. Experiments following the theory proposals confirmed the KZM scaling with very good accuracy~\cite{ulm2013observation,pyka2013topological,mielenz2013trapping}. 

\subsubsection{Influence of dissipation/decoherence}
For open quantum systems the focus has been on how decoherence/dissipation affects the creation of defects during the quench. The rule of thumb is that fluctuations stemming from any environment will increase the amount of defects~\cite{fubini2007robustness,cincio2009dynamics,dutta2016anti}. For the random Ising model, the characteristics may also be qualitatively different from those predicted by the KZM for closed systems~\cite{dziarmaga2006dynamics}. In particular, even at infinitely slow quenches the density of defects is non-vanishing also for finite systems where the gap is always non-zero. In Refs.~\cite{fubini2007robustness,dutta2016anti}, studying a system belonging to Class II in the classification above, it was especially found that as the quench became slower the number of excitations increased, something that was termed `anti-Kibble-Zurek mechanism'. Such a behavior is explained from the fact that a slower process implies a longer duration for which the system interacts with its environment, such that the environment induced excitations become essential. The same phenomenon is known both in `quantum coherent control'~\cite{ivanov2004effect,mathisen2016view} and adiabatic quantum computing~\cite{sarandy2005adiabatic}, i.e. there is an optimal process time for minimizing the amount of excitations.

\subsubsection{Preparation of NESS's }
Using environments and dissipation as resources have been especially discussed in the realm of quantum information processing. The Liouvillian gap (\ref{lgap}) plays the role for open systems as what the energy gap (above the ground state) does for Hamiltonians. For a large gap, the steady state or ground state is robust to external fluctuations. However, steady states are not only protected against external imperfections due to the large size of the gap, the gap also implies that if you are taken out of the steady state manifold you will relax back to it. If your desired steady state is unique you are guaranteed to be projected back onto it if the disturbances are temporary.   

For quantum information processing the natural candidate for steady states are non-classical entangled states. Several proposals have been put forward how dissipation can be harnessed for preparation of entangled states, e.g. Bell states~\cite{clark2003unconditional,kraus2004discrete,paternostro2004complete,cho2011optical,kastoryano2011dissipative}. It has also been experimentally demonstrated when it comes to the creation of Bell states of trapped ions, either via discrete gates~\cite{barreiro2011open} or continuous relaxation~\cite{lin2013dissipative}, or to entanglement between macroscopic atomic clouds~\cite{krauter2011entanglement}. Extending the schemes for the generation of multi-qubit entangled states have also been suggested~\cite{kraus2008preparation} and experimentally verified~\cite{barreiro2010experimental,schindler2012quantum}, or other exotic many-body quantum states~\cite{diehl2008quantum,witthaut2008dissipation,diehl2010dynamical,diehl2011topology,muller2012engineered,bardyn2013topology}. 

As discussed already in the previous Subsection, in driven systems dissipation may cause critical behavior of the steady states~\cite{diehl2008quantum,diehl2010dynamical,eisert2010noise,hoening2012critical,sieberer2013dynamical,eisert2014quantum,hannukainen2017dissipation}. Criticality in driven-dissipative systems has recently been experimentally explored in quantum optical systems like the Dicke `normal-superradiance' PT~\cite{baumann2009dicke,baumann2011exploring,klinder2015dynamical}. Other examples of steady state criticality are `optical bistability'~\cite{bonifacio1978optical,drummond1980quantum,fink2017observation} and the onset of lasing in the laser~\cite{degiorgio1970analogy,rice1994photon,haken2012laser}.


\section{General description of the Kibble-Zurek mechanism for NESS phase transitions}\label{sec3}

\subsection{The Bloch representation for Lindblad master equations}\label{ssec3A}
The KZM relies on the AI-approximation, and while it is clear what adiabatic vs. diabatic (impulse) evolutions mean for closed quantum systems it is not completely evident what it implies for open quantum systems. We therefor must define what we mean by adiabaticity for open quantum systems. As we will see there are various approaches one may take, but before that we need to say something about the Liouvillian of Eq.~(\ref{lindblad}).

The Lindblad master equation~(\ref{lindblad}) is linear, i.e. $\hat{\mathcal{L}}(c_1\hat\rho_1+c_2\hat\rho_2)=c_1\hat{\mathcal{L}}(\hat\rho_1)+c_2\hat{\mathcal{L}}(\hat\rho_2)$. If we parametrize the density operator $\hat\rho$ and represent it as some vector, the linearity implies that the Lindblad equation is cast in a simple matrix form. There are different choices for how to parametrize $\hat\rho$, but here we employ one related to the Bloch vector representation. Given that the Hilbert space dimension is $D$, any density operator obeying positivity and normalization can be expressed as~\cite{kimura2003bloch,bertlmann2008bloch}
\begin{equation}\label{dens}
\hat\rho=\frac{1}{D}\left(\mathbb{I}+\sqrt{\frac{D(D-1)}{2}}\mathbf{R}\cdot\lambda\right).
\end{equation}
Here, $\mathbf{R}=(r_1,\,r_2,\,\dots,\,r_{D^2-1})$ is the generalized (real) Bloch vector and the vector $\lambda=(\hat\lambda_1,\,\hat\lambda_2,\dots,\,\hat\lambda_{D^2-1})$ is composed of the generalized Gell-Mann matrices $\hat\lambda_i$\cite{hioe1981n}. The $\hat\lambda_i$ matrices are generators of the Lie algebra corresponding to the group $SU(D)$, and in particular they are traceless, hermitian, and mutually orthogonal such that given any density operator its Bloch vector is obtained from $r_i=\mathrm{Tr}[\hat\lambda_i\hat\rho]$. For $D=2$ (qubit) and $D=3$ (qutritt) the matrices are the standard Pauli and Gell-Mann matrices respectively. The Bloch vector length $|\mathbf{R}|\leq1$ such that the state space can be represented by a `Bloch hyper-sphere'. However, whenever $D>2$ not all points in the Bloch sphere represent a physical state~\cite{kimura2003bloch,goyal2016geometry}. The larger dimension, the sparser the Bloch sphere is in terms of physical states. The non-physical states are not proper density matrices as they are not positive semi-definite.  

In the Bloch representation we have parametrized the density operator in terms of $\mathbf{R}$, and the Lindblad master equation is given by~\cite{schirmer2010stabilizing}
\begin{equation}\label{meq}
\partial_t\mathbf{R}=\mathbf{MR}+\mathbf{b}.
\end{equation}
The Liouvillian matrix $\mathbf{M}$ is of dimension $(D^2-1)\times(D^2-1)$ and in general not hermitian. In fact, for a closed system $\mathbf{M}$ is skew-symmetric. For now, let us assume that both $\mathbf {M}$ and $\mathbf{b}$ are time-independent. The term $\mathbf{b}$ is a column vector of $D^2-1$ elements and represents some sort of pumping that prevents the state $\mathbf{R}={\bf 0}$ to be a trivial steady state. The general steady state is given by $\mathbf{MR}_\mathrm{ss}+\mathbf{b}=0$, or if $\mathbf{M}$ is invertible $\mathbf{R}_\mathrm{ss}=-\mathbf{M}^{-1}\mathbf{b}$. If $\mathbf{M}$ is not invertible the system of equations is under-determined and the steady state need not be unique which will give more interesting situations when discussing the KZM for NESS critical models. When $\mathbf b=0$ and $\mathbf M$ is not invertible we note that the steady state $\mathbf M\mathbf R_\mathrm{ss}=0$ defines a connected manifold of steady states due to the ambiguity of the norm of $\mathbf R_\mathrm{ss}$. The continuity and linearity of (\ref{meq}) warrant that there must exist at least one steady state (or fixed point)~\cite{schirmer2010stabilizing}. In the general case of a non-vanishing pump term $\mathbf{b}$, by solving the homogeneous equation $\partial_t\mathbf{Q}=\mathbf{MQ}$ the solution of the inhomogeneous problem is $\mathbf{R}(t)=\mathbf{Q}(t)\left[\mathbf{R}(0)+\int_0^td\tau\,\mathbf{Q}^{-1}(\tau)\mathbf{b}(\tau)\right]$. Or if we introduce the matrices $\mathbf{V}$ and $\mathbf{U}$ that diagonalizes the Liouvillian matrix, i.e. $\mathbf {D}=\mathbf{V}^t\mathbf{MU}$ with $\mathbf{D}$ diagonal, then the right eigenvectors of $\mathbf{M}$ evolve as $\mathbf{Q}_i(t)=\mathbf{Q}_i(0)e^{\mu_it}+(e^{\mu_it}-1)\mathbf{V}^t\mathbf{b}/\mu_i$ where $\mu_i$ is the corresponding eigenvalue of $\mathbf{M}$. Thus, the eigenvalues of the Liouvillian matrix determines the characteristic time-scales, and note that we must have $\mathrm{Re}(\mu_i)\leq0$ in order to preserve normalization. The Liouvillian gap is defined as before in Eq.~(\ref{lgap}). Provided that the Liouvillian matrix is diagonalizable (for example if it is normal, $[\mathbf{M},\mathbf{M}^\dagger]=0$, as will be relevant for us in the following Section), its eigenvectors $\mathbf{Q}_i$ (or $\mathbf{R}_i$ if $\mathbf{b}={\bf 0}$) of $\mathbf{M}$ form an over-complete basis, i.e. they are not linearly independent. Furthermore, given one eigenvector, its corresponding density operator $\hat\rho_i$ need not be physical. Nevertheless, any initial physical $\mathbf{R}(0)$ will evolve into a new physical Bloch vector $\mathbf{R}(t)$.  

The fact that $\mathbf{M}$ is not hermition implies that it might not be diagonalizable. There exists, however, a similarity matrix $\mathbf{S}$ that puts $\mathbf{M}$ on a Jordan block form, i.e.  
\begin{equation}
\mathbf{D}=\mathbf{SMS}^{-1},
\end{equation}
where the Jordan blocks of $\mathbf{D}$ have identical values on the diagonal and ones on the superdiagonal, and the remaining values are all zero. Thus, if $\mathbf{M}$ is diagonalizable all Jordan blocks has dimension one. If $\mathbf{M}$ is not diagonalizable, the spectrum shows an exceptional point~\cite{heiss2004exceptional,heiss2012physics}, where the real parts of at least two eigenvalues of $\mathbf{M}$ coalesce. Furthermore, at the exceptional point the corresponding eigenvectors are identical. The presence of exceptional points will however not be relevant for us in the examples discussed in the following section.

\subsection{Adiabaticity for open quantum systems}\label{adsubsec}
Having introduced the Bloch representation, it is rather straightforward to generalize the ideas of adiabaticity from quantum mechanics~\cite{ballentine2014quantum} to open quantum systems~\cite{sarandy2005adiabatic2,sarandy2005adiabatic}. In order to make the analysis transparent we here consider the cases with vanishing pump terms (the generalization to $\mathbf{b}\neq0$ is direct), and we call the instantaneous right eigenvectors of $\mathbf {M}(t)$ for $\mathbf {R}_i(t)$. Adiabaticity then implies that the dynamics does not generate any transfer of population between the instantaneous eigenvectors. Thus, if we initialize the system in one eigenvector $\mathbf{R}_i(0)$, and the system evolves adiabatically the state at a later time is $\mathbf{R}(t)=\mathbf{R}_i(t)\exp\left(\int_0^t\mu_i(\tau)\,d\tau\right)$. The condition warranting adiabatic evolution for an open quantum system described by a Liouvillian matrix $\mathbf{M}(t)$ takes a very similar form to that for Hamiltonian systems~\cite{sarandy2005adiabatic2}. That is, if $\omega_{ij}(t)=|\mu_i(t)-\mu_j(t)|$ represents the ``gap'' and $\mathbf{L}_j(t)$ a left eigenvector of $\mathbf{M}(t)$, then the `rate of change' $|\mathbf{L}_j(t)\dot{\mathbf{M}}(t)\mathbf{R}_i(t)|$, with dot representing time derivative, should be small in comparison to the ``gap'' $\omega_{ij}^2(t)$ for all times and all $j$.

The similarities between adiabatic evolution generated by a hermitian matrix $\mathbf{H}(t)$ and a non-hermitian matrix $\mathbf{M}(t)$ are many, but there are still important differences that should be appreciated; ($i$) we already mentioned that the set of eigenvectors $\mathbf{R}_i$ forms an over-complete basis, ($ii$) most of the eigenvectors do not represent physical states $\hat\rho_i$, ($iii$) contrary to state vectors, the norms of the Bloch vectors are arbitrary between 0 and 1, and ($iv$) since $\mathrm{Re}(\mu_i)\leq0$ even under adiabatic evolution the norm of the adiabatic Bloch vector typically decreases. If the instantaneous eigenstates $\mathbf{R}_i(t)$ define the adiabatic states $\hat\rho_i^{\mathrm{(ad)}}(t)$, the ambiguity of the Bloch vector norm means that for every $i$ there is a set comprised of infinitely many adiabatic states. The last point in the list above then implies that even under adiabatic evolution an adiabatic state evolve within the $i$'th set of adiabatic states.

The generalization of the adiabatic theorem, going from a Hamiltonian $\hat H(t)$ to a non-hermitian matrix $\mathbf{M}(t)$ as sketched above, is natural from a mathematical perspective. It provides, however, a less clear physical picture, especially since most eigenvectors $\mathbf{R}_i$ are non-physical. Naturally, we are more interested in how physical states evolve, and what is meant by adiabaticity for those. The physical  state of special interest is the instantaneous steady state
\begin{equation}
\hat{\mathcal L}_t(\hat\rho_\mathrm{ss}^\mathrm{(ad)}(t))=0,
\end{equation}
where the subscript $t$ marks that the Liouvillian is explicitly time-dependent and the superscript (ad) denotes an {\it adiabatic} steady state. Thus, $\hat\rho_\mathrm{ss}^\mathrm{(ad)}(t)$ is an exceptional example of the adiabatic states $\hat\rho_i^{\mathrm{(ad)}}(t)$ introduced in the previous paragraph. Note that if we imagine a fixed time $t$, $\hat\rho_\mathrm{ss}^\mathrm{(ad)}(t)$ is a corresponding steady state. If the time-dependence of the Liouvillian is weak we may expect that the relaxation time of the system is short on the overall time-scale such that throughout the driving the system stays close to a steady state. This then resemblance adiabatic evolution.

Analyzing adiabaticity in terms of deviations from the instantaneous steady state has been the subject of several papers~\cite{davies1978open,joye2007general,avron2012adiabatic,venuti2016adiabaticity}. In particular, if the quench can be assigned a time $T$ the deviations from $\hat\rho_\mathrm{ss}^\mathrm{(ad)}(t)$ deep in the adiabatic regime scales as $T^{-\eta}$, with $\eta=1$ for a gapped system and for a system in which the gap closes $\eta=\frac{1}{1+\beta}$, where the exponent $\beta$ determines how fast the Liouvillian gap (\ref{lgap}) closes~\cite{venuti2016adiabaticity}. Since $\beta\leq0$, any gap closening worsen the adiabaticity. $\beta$ should be compared to $\nu z$, and $T$ to $\tau_Q$ for the closed system discussed in Sec.~\ref{ssec:KZM}.

\begin{figure}
\includegraphics[width=8cm]{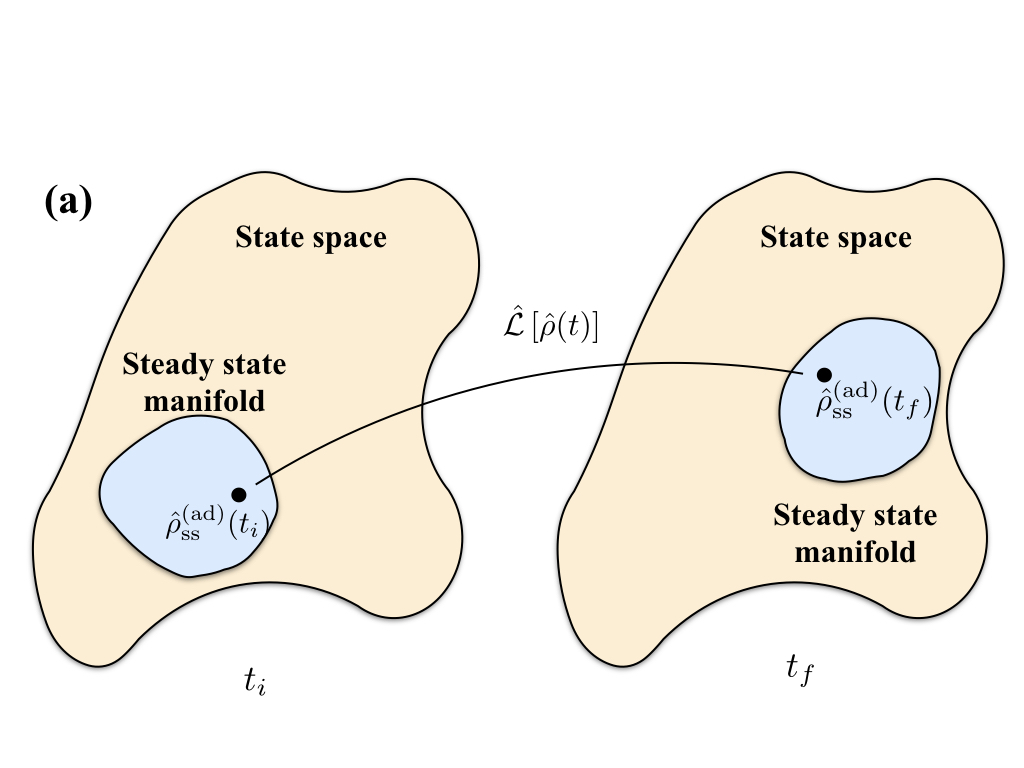} 
\includegraphics[width=8cm]{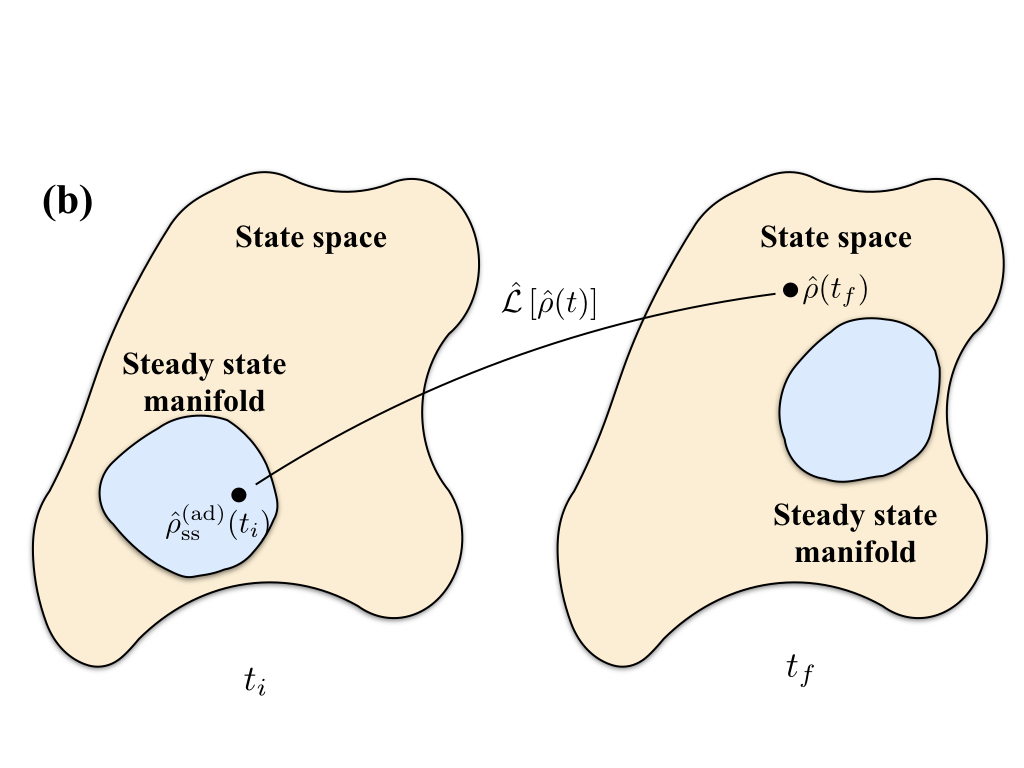} 
\includegraphics[width=8cm]{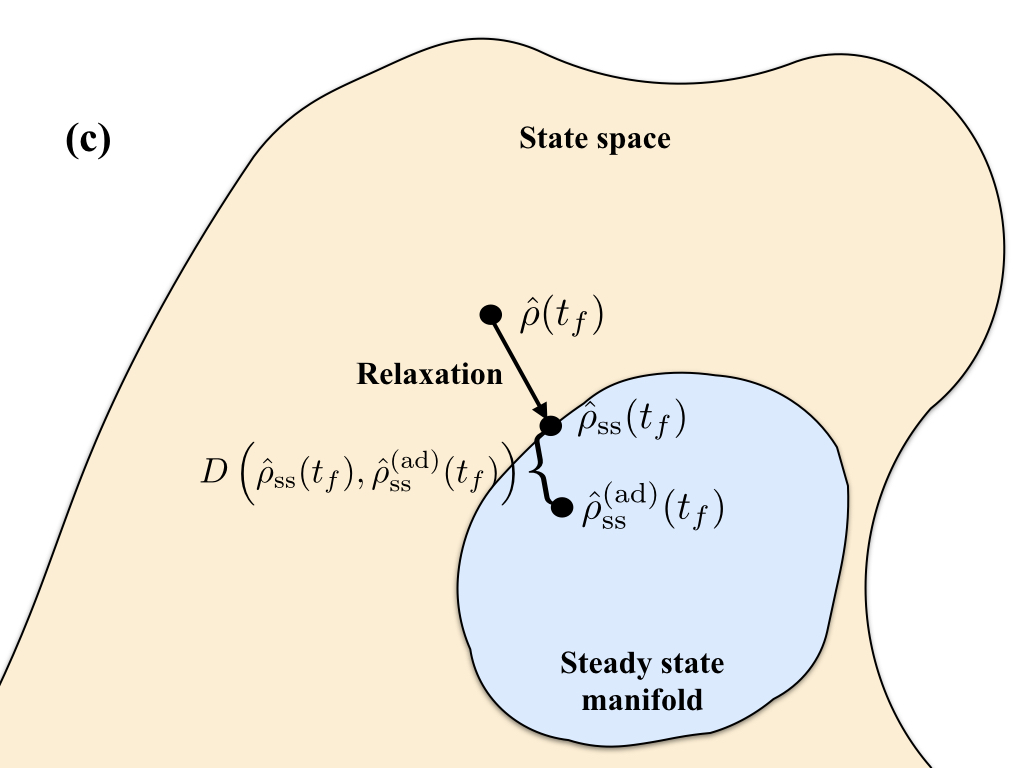} 
\caption{Schematic picture of the time-evolution for open quantum systems. The yellow area represents the space of all physical states $\hat\rho(t)$, and the light blue the manifold of instantaneous steady states $\hat\rho_\mathrm{ss}(t)$. Since the model is explicitly time-dependent, the steady state manifold could change in time. In (a) the evolution is adiabatic and an initial steady state $\hat\rho_\mathrm{ss}^\mathrm{(ad)}(t_i)$ is adiabatically propagated by the Liouvillian $\hat{\mathcal L}_t$ to a final steady state $\hat\rho_\mathrm{ss}^\mathrm{(ad)}(t_f)$. The steady state manifold is continuously deformed as time progresses and the propagated state $\hat\rho_\mathrm{ss}^\mathrm{(ad)}(t)$ remains in the manifold throughout. There is a one-to-one mapping between every steady states in the different manifolds at different times. For a non-adiabatic evolution, the initial steady state is taken out from its instantaneous steady state into some state $\hat\rho(t)$, typically lying outside the steady state manifold, as depicted in (b). When the driving stops at time $t_f$, the state $\hat\rho(t)$ relaxes down to a steady state $\hat\rho_\mathrm{ss}(t_f)$ in the steady state manifold (c). The distance $D$ is then a measure of how non-adiabatic the whole process was.
}
\label{fig3}
\end{figure}
 
\subsection{Measure of non-adiabatic excitations}\label{ssec3C}
The KZM for quantum systems estimates the density of defects in terms of the quench rate $\tau_Q$ and the critical exponents according to Eq.~(\ref{defdens}). These defects are topological by nature and describe local excitations. Thus, the sum of them gives the amount of excitations above the ground state energy. The more excited the system gets, while quenched through the critical point, the higher number of defects. 

When we deal with an open system there is nothing like the ground state energy. We can of course consider the instantaneous energy $E_H(t)=\mathrm{Tr}[\hat\rho(t)\hat H(t)]$ where $\hat\rho(t)$ is the state at time $t$. This energy could be compared to the `adiabatic' energy $E_H^\mathrm{(ad)}(t)=\mathrm{Tr}[\hat\rho_\mathrm{ss}^\mathrm{(ad)}(t)\hat H(t)]$ to give some sort of `excitation measure'. However, such a comparison only makes sense if the influence of the environment is modest and the full system is mainly prescribed by the Hamiltonian. In fact, previous works have had this idea in mind; how is the quench affected by a weak coupling to an environment~\cite{dziarmaga2006dynamics,fubini2007robustness,cincio2009dynamics,nalbach2015quantum,dutta2016anti}? The general answer to this question is that the environment causes additional fluctuations that increase the amount of excitations. When the system plus environment cannot be thought of as separate subsystems, for example when the criticality crucially depends on both parts, $E_H(t)-E_H^\mathrm{(ad)}(t)$ has little to do with any excitations. Learning from the discussion above about adiabaticity in open quantum systems, the proper measure for adiabaticity should be the distance from the instantaneous steady state $\hat\rho_\mathrm{ss}^\mathrm{(ad)}(t)$. Indeed, $\hat\rho_\mathrm{ss}^\mathrm{(ad)}(t)$ is the state of relevance and any non-adiabatic excitations should be measured relative to this state. There are many possible choices for such a distance, but as we will see the natural one here is the trace distance~\cite{nielsen2010quantum}, which for two density matrices is defined as
\begin{equation}\label{tr1}
D(\hat\rho_1,\hat\rho_2)\equiv\frac{1}{2}\mathrm{Tr}\left[\sqrt{(\hat\rho_1-\hat\rho_2)^2}\right]=\frac{1}{2}\sum_i|\lambda_i|,
\end{equation}
where $\lambda_i$ is the $i$'th eigenvalue of $\hat\rho_1-\hat\rho_2$. Thus, the quantity of interest for us is $D\left(\hat\rho(t),\hat\rho_\mathrm{ss}^\mathrm{(ad)}(t)\right)$. The idea behind the trace distance as the appropriate measure is pictured in Fig.~\ref{fig3}, and we will visualize it further in the following section when we discuss particular examples.

Since any steady state has eigenvalue zero of the Liouvillian $\hat{\mathcal L}_t$, under adiabatic evolution the Liouvillian does not generate any evolution of $\hat\rho_\mathrm{ss}^\mathrm{(ad)}(t)$. That is, given that we started in the instantaneous steady state $\hat\rho_\mathrm{ss}^\mathrm{(ad)}(0)$, then $D\left(\hat\rho(t),\hat\rho_\mathrm{ss}^\mathrm{(ad)}(t)\right)=0$ for all times. In particular, the length of the corresponding Bloch vector is constant. Any non-zero trace distance can only be the result of non-adiabatic transitions ocurring during the evolution. Typically, transitions out from $\hat\rho_\mathrm{ss}^\mathrm{(ad)}(t)$ will put the system in a non-steady state. The real parts of the Liouvillian eigenvalues will push the state back towards the steady state manifold, simultaneously as the driving may cause further non-adiabatic excitations. As the driving stops, according to the discussion above regarding fixed points of (\ref{meq}), the relaxation maintains and eventually the system is to be found in a steady state. The distance of this steady state to the adiabatic steady state, 
\begin{equation}\label{trdist}
D\left(\hat\rho_\mathrm{ss},\hat\rho_\mathrm{ss}^\mathrm{(ad)}\right),
\end{equation} 
is a direct result from non-adiabatic transitions developed during the quench and hence serves as our measure of excitations. This is one of the key results of the present work.

We typically consider a situation where the final times are far from the impulse regime. For the quench in the closed system this implies that the evolution is adiabatic and no further excitations develop. For the open case, during the later stages far from the critical point when the system relaxes to its instantaneous steady state -- the relaxation time-scale is typically the short one. Thus, we normally have that $\hat\rho(t_f)$ is to a good approximation an instantaneous steady state as long as $t_f$ is large in comparison to any freeze-out time. The relaxation depicted in Fig.~\ref{fig3} (c) therefor occurs already before we reached the final time $t_f$. With this in mind it should be clear that the trace distance is not only the suitable measure in terms of quantifying the amount of non-adiabatic excitations, but it also has the advantage that it does not depend on $t_f$ as long as it is much larger than the freeze-out time $\hat t$. This observation is important since our aim is to describe universal features of the quench for open quantum critical systems. As a universal quantity we do not want it to depend on $t_i$ nor $t_f$. This is different from earlier works exploring quenches through critical points in open quantum systems~\cite{dziarmaga2006dynamics,fubini2007robustness,cincio2009dynamics,nalbach2015quantum,dutta2016anti}. 

In the absence of an environment we wish that the trace distance (\ref{trdist}) should in some sense be directly related to the amount of non-adiabatic excitations out from the ground state. As we already pointed out, in certain situations one cannot even talk about energy excitations since there is no energy spectrum. However, in the cases when one can it is of course desirable that the trace distance measures such energy excitations. Class I of the above classification is such a scenario. Here   criticality cannot arise from coupling to the environment since the jump operator commutes with the Hamiltonian -- the environment induces a dephasing in the energy basis. In particular, in this Class all steady states are diagonal in the energy eigenbasis. The instantaneous ground state of $\hat H(t)$ is thereby also an instantaneous steady state. With $|E_n(t)\rangle$ the instantaneous (adiabatic) eigenstates we envision the situation where the system is initialized in the Hamiltonian ground state, and hence for adiabatic evolution $\hat\rho_\mathrm{ss}^\mathrm{(ad)}(t)=|E_0(t)\rangle\langle E_0(t)|$. The actual time-evolved state, after relaxation to its final steady state, is on the form $\hat\rho_\mathrm{ss}(t)=\sum_{n=0}^\infty p_n|E_n(t)\rangle\langle E_n(t)|$. The trace distance at the final time $t_f$ becomes
\begin{equation}\label{trdist2}
D\left(\hat\rho_\mathrm{ss}(t_f),\hat\rho_\mathrm{ss}^\mathrm{(ad)}(t_f)\right)=\sum_{n=1}^\infty p_n=1-p_0.
\end{equation} 
Thus, the distance gives the probability to be excited. Furthermore, if we shift the instantaneous ground state energy $E_0(t_f)=0$, the amount of excitations in terms of energy is simply $\delta E=\mathrm{Tr}\left[\hat\rho_\mathrm{ss}(t_f)\hat H(t_f)\right]$. 

\subsection{The KZM for open quantum systems}\label{ssec3d} 
As we discussed in Sec.~\ref{adsubsec}, the adiabatic concepts for time-dependent hermitian matrices (i.e. Hamiltonians) can be generalized to in principle any square matrix like for example the Liouvillian matrix $\mathbf M(t)$. Regardles of the properties of the matrix the idea is that the variations of the change should be small in comparison to the squared gap in order to warrant adiabatic evolution. Since the eigenvalues $\mu_i(t)$ of $\mathbf M(t)$ are in general complex, the absolute value of the gap should be considered, i.e. $\omega_{ij}(t)=|\mu_i(t)-\mu_j(t)|$. Remember that even if the pump term $\mathbf b(t)$ is non-zero, a transformation can cast the Bloch equation into a homogenous one such that we can limit the discussion to homogenous equations.

We already pointed out that there is, however, one crucial aspects of adiabaticity in open quantum systems differing from the standard setting of closed quantum systems. As long as the adiabatic state $\mathbf R_i(t)$ is not a steady state, even under adiabatic evolution the norm of $\mathbf R_i(t)$ will decrease (keep in mind that we assume $\mathbf b(t)=0$, and if this is not the case it is the norm of $\mathbf Q_i(t)$ that is shrinking). Thus, on the Bloch sphere any symmetry axis defines a class of connected adiabatic states. 

As for QPT's, criticality of open quantum systems is also accompanied by a gap closening in the spectrum of $\mathbf M$~\cite{kessler2012dissipative}. Thus, at the critical point the Liouvillian gap of Eq.~(\ref{lgap}) $\Delta_M\rightarrow0$. This is the equivalence of the critical slowing down for open quantum PT's. Far from the critical point we expect instead that $\omega_{ij}(t)$, for any $i$ and $j$, is large compared to the inverse rate-of-change, such that the dynamics is adiabatic. In this respect we are lead to introduce the AI-approximation. The picture is then very similar to that of Fig.~\ref{fig1}, where the evolution is divided into an adiabatic, an impulse, and an adiabatic regime. Recall that the freeze-out time $-\hat t$ when the system goes from adiabatic to diabatic is  determined from equalizing the relaxation time $\tau_H=\Delta_H^{-1}$ with the inverse transition rate. For a linear quench we thereby have $\tau_H(\hat t)=\alpha\hat t$, where $\alpha$ is some parameter that could depend on the system parameters~\cite{damski2006adiabatic}. When considering a quench through a critical point of an open quantum system we should replace the Hamiltonian reaction time $\tau_H$ with the Liouvillian reaction time $\tau_M=\omega_{ij}^{-1}$ (note that the Liouvillian gap $\Delta_M$ sets the relaxation rate to the steady state, while $\omega_{ij}$ determines total response), and furthermore the parameter $\alpha$ may well be altered by the environment and especially is expected to depend on the loss rates $\kappa_i$. In fact, it is not {\it a priori} clear that the characteristic time for the rate-of-change will be linear even though the quench is linear. One could, for example, imagine Lindblad jump operators that are explicitly time-dependent. Another possibility is that the impulse regime need not be symmetric around the critical point such that there is a left $\hat t_L$ and right $\hat t_R$ freeze-out time with $\hat t_L\neq-\hat t_R$. Nevertheless, with the above argument we hope that it is cleatr that the general idea of the AI-approximation can be applied to critical open quantum systems. In the following section we will verify this by analyzing two examples.


\section{Examples}\label{sec4}
For a non-trivial situation we need a model that supports a manifold of many steady states. For a single unique steady state it is clear that by taking $t_f$ large enough the system always ends up in this state and we cannot conclude how adiabatic the quench was. We know that Class I of our classification supports a connected manifold of steady states. This Class is also physically relevant as it describes energy dephasing. In fact, small fluctuations in experimental parameters should at first cause a dephasing in the energy basis. We thereby look for Lindblad jump operators that commute with the Hamiltonian $\hat H(t)$ for all times. One choice is to take the instantaneous projectors onto the adiabatic states; $\hat L_i(t)=|\phi_i^{\mathrm{(ad)}}(t)\rangle\langle \phi_i^\mathrm{(ad)}(t)|$. Alternatively one can take that $\hat L(t)=\hat H(t)$. In both examples we have that the jump operators are explicitly time-dependent and in a strict sense the resulting master equation is not on a Lindblad form. Nevertheless, this is not so important for us since  it is still a `completely positive trace preserving map' that guarantees that the density matrix stays physical under time-evolution. In both examples discussed in this section we consider the case when the jump operators equals the system Hamiltonian. Using the adiabatic state projectors instead does not change the results qualitatively in any way. 

\subsection{Dephasing LZ model}\label{ssec4A}
\begin{figure}
\includegraphics[width=8cm]{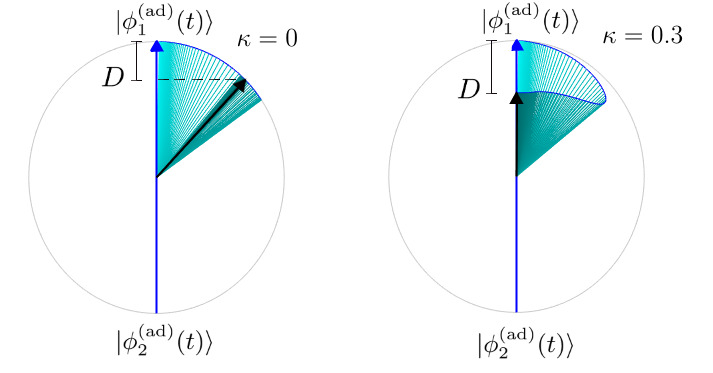} 
\caption{(Color online) Comparison between the Bloch vector evolution for the closed, $\kappa=0$, and open, $\kappa\neq0$, LZ model. The thick blue arrow represents the manifold of steady states (symmetry axis); the arrow head is the adiabatic state $|\phi_1^\mathrm{(ad)}(t)\rangle$ and the other end of the blue arrow is the orthogonal adiabatic state $|\phi_2^\mathrm{(ad)}(t)\rangle$. The thick black arrow shows the final state at $t_f$ (big enough such that the relaxation is complete). Adiabatic evolution would mean that the black and blue arrows completely overlap. The turquoise thin lines give snapshots of the Bloch vector $\mathbf R(t_i)$ at different times $t_i$. The decreased Bloch vector for $\kappa\neq0$ is evident, as is the relaxation down to the steady state manifold. The amount of excitations is determined from $D$. For this numerical example, $v=0.4$, $g=0.5$, and the initial and final times $t_i=-10$ and $t_f=10$ are taken to warrant convergence of the population transfer.}
\label{fig4}
\end{figure}

As Damski pointed out, the simplest model supporting the KZM is the LZ problem~\cite{damski2005simplest}. It is not a model describing true criticality, but rather a smooth transition between two orthogonal states, see Fig.~\ref{fig2}. For the dephasing LZ problem the Lindblad equation takes the form
\begin{equation}\label{lzlind}
\begin{array}{l}
\partial_t\hat\rho(t)=i\left[\hat\rho(t),\hat H_\mathrm{LZ}(t)\right]\\ \\ 
\displaystyle{+\kappa\left(2\hat H_\mathrm{LZ}(t)\hat\rho(t)\hat H_\mathrm{LZ}(t)-\hat H_\mathrm{LZ}^2(t)\hat\rho(t)-\hat\rho(t)\hat H_\mathrm{LZ}^2(t)\right),}
\end{array}
\end{equation}
with $\hat H_\mathrm{LZ}(t)$ the LZ Hamiltonian~(\ref{lzham}). The manifold of steady states comprises those along a symmetry axis in the Bloch sphere between the adiabatic states, $|\phi_1^\mathrm{(ad)}(t)\rangle$ and $|\phi_2^\mathrm{(ad)}(t)\rangle$. For large negative or positive times these states approximately coincide with the diabatic states $|0\rangle=[1\,\,0]^T$ and $|1\rangle=[0\,\, 1]^T$, i.e. the north and the south pole on the sphere. Starting in say the south pole, regardless of whether $\kappa$ vanishes or not, adiabatic evolution implies that the state stays pure and traverses the Bloch sphere and ends up on the north pole. In the adiabatic basis, the state is frozen under adiabatic evolution. Non-adiabatic excitations takes the state away from the symmetry axis. Simultaneously, the dephasing shrinks the length of the Bloch vector and pushes the state back towards the symmetry axis (steady state manifold). This relaxation is clearly absent when $\kappa=0$. The result of a numerical simulation demonstrating this behavior is presented in Fig.~\ref{fig4}.

\begin{figure}
\includegraphics[width=8cm]{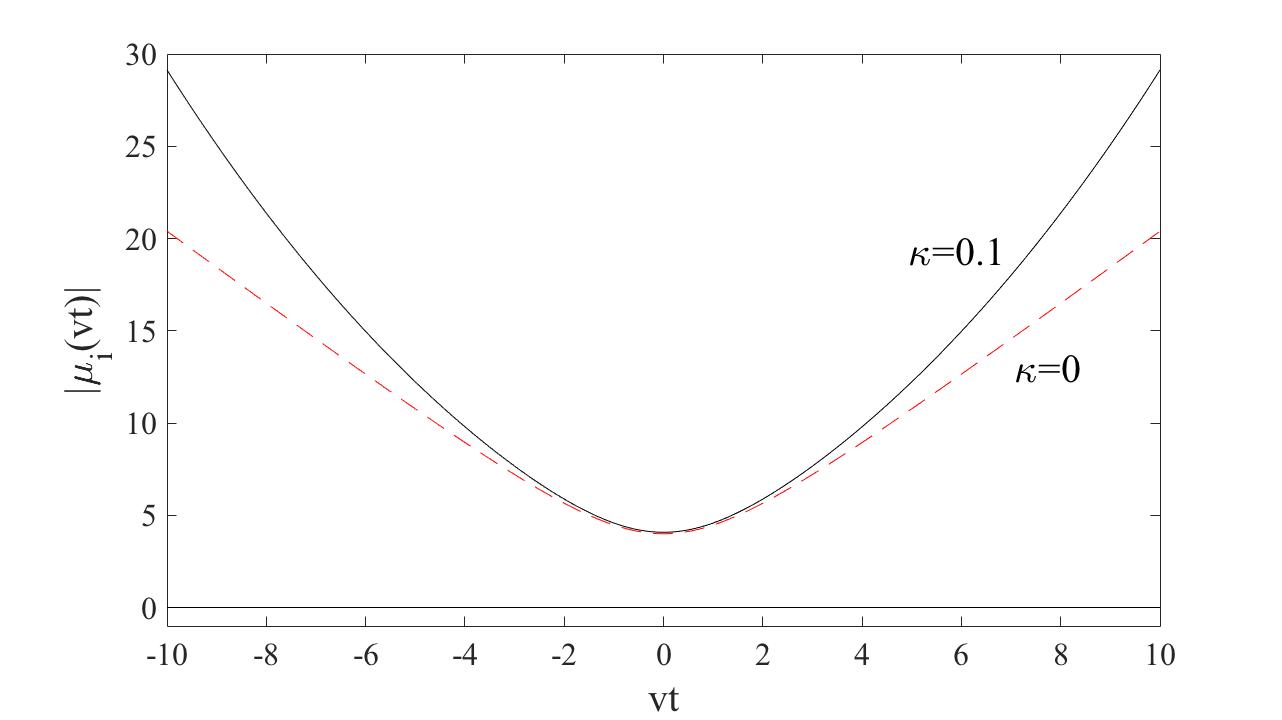} 
\caption{(Color online) Absolute values of the instantaneous eigenvalues (\ref{meig}) of the Liouvillian matrix (\ref{mmatrix}). There is a time-independent zero eigenvalue corresponding to the steady states. 
The difference between the zero eigenvalue and the remaining two reflects the spectral gap function $\omega(t)$ which sets the inverse response time. In the figure $g=2$ and we show two examples; $\kappa=0.1$ for the solid black line and $\kappa=0$ for the dashed red line. Note, in particular, that the gap $\omega(t)$ increases with $\kappa$ implying that the response time gets shorter the larger $\kappa$ is.}
\label{fig5}
\end{figure}

In Appendix~\ref{appA} we give the general expression for the Bloch equation for two-level systems, from which we can extract the corresponding equations for (\ref{lzlind}). The Liouvillian matrix takes the form
\begin{equation}\label{mmatrix}
\mathbf{M}(t)=2\left[
\begin{array}{ccc}
-\kappa (vt)^2 & -vt & \kappa gvt\\
vt & -\kappa\left[(vt)^2+g^2\right] & -g\\
\kappa gvt & g & -\kappa g^2
\end{array}\right]
\end{equation}
and the pump term $\mathbf b=0$, as follows from that the jump operator is hermitian. The Liouvillian matrix is normal, $\left[\mathbf M(t),\mathbf M^\dagger(t)\right]=0$, which implies that it is unitarilly diagonalizable~\cite{arfken2005mathematical}. The instantaneous eigenvalues are
\begin{equation}\label{meig}
\begin{array}{l}
\mu_1(t)=0,\\ \\
\mu_2(t)=-2\left(\kappa\varepsilon^2(t)-i\varepsilon(t)\right),\\ \\
\mu_3(t)=\mu_2^*(t)=-2\left(\kappa\varepsilon^2(t)+i\varepsilon(t)\right),
\end{array}
\end{equation}
where, as before, $\varepsilon(t)=\sqrt{(vt)^2+g^2}$. The corresponding, orthonormal, instantaneous eigenvectors are
\begin{equation}
\begin{array}{c}
\begin{array}{ll}
\mathbf R_1(t)=\frac{1}{\varepsilon(t)}\left[
\begin{array}{c} g\\ 0\\ vt\end{array}\right], & 
\mathbf R_2(t)=\frac{1}{\sqrt{2}\varepsilon(t)}\left[
\begin{array}{c} vt\\ -i\varepsilon(t) \\ g\end{array}\right],
\end{array}\\ \\
\begin{array}{c} 
\mathbf R_3(t)=\frac{1}{\sqrt{2}\varepsilon(t)}\left[
\begin{array}{c} vt\\ i\varepsilon(t) \\ g\end{array}\right].
\end{array}
\end{array}
\end{equation}
The first Bloch vector $\mathbf R_1(t)$ is the steady state with the accompanying zero eigenvalue. The remaining two Bloch vectors are both complex, and hence cannot represent physical states. The spectral gap $\omega(t)=|\mu_1(t)-\mu_2(t)|=|\mu_1(t)-\mu_3(t)|=2\sqrt{\kappa^2\varepsilon^4(t)+\varepsilon^2(t)}$. For $\kappa=0$ we regain the LZ gap $2\varepsilon(t)$. The absolute values of the eigenvalues are shown in Fig.~\ref{fig5}, from where it is also evident that the gap grows with $\kappa$.

Turning to the KZM for the open LZ problem, we note that the response time $\tau_M=\omega^{-1}(t)$ is a decreasing function of $\kappa$. This suggests that the non-vanishing loss rate makes the quench more adiabatic. In return, this should result in a shorter impulse region, i.e. the value of the freeze-out time $|\hat t|$ decreases. But this is in contrast to the general knowledge that the coupling to an environment causes further excitations stemming from the additional fluctuations induced by the environment~\cite{fubini2007robustness,cincio2009dynamics,dutta2016anti}. Indeed, one finds that the amount of excitations do increase for a non-zero $\kappa$ (see below). Does this mean that the idea of the KZM, as a result of an AI approximation, breaks down for open systems? As pointed at in Sec.~\ref{ssec3d}, the answer to this question lies in the fact that the rate-of-change also depends on $\kappa$, i.e. the slope of the red dashed line in Fig.~\ref{fig1} is $\kappa$-dependent. Thus, even if the solid black lines of Fig.~\ref{fig1} move downward (suggesting a shorter impulse region), the change in the slope of the red dashed line can compensate this to keep the impulse region large. If the amount of excitations increases we even expect the impulse region to grow with $\kappa$.

For the closed LZ problem it was motivated that the slope of the red dashed line in Fig.~\ref{fig1} is given by $\alpha=\pi$~\cite{damski2006adiabatic}. In particular, in the regime of relatively fast quenches where we expect the AI-approximation to be applicable, an expansion argument, and comparing to the known analytical result~(\ref{lzformula}), give this slope value. To analyze the open LZ problem we rewrite the Liouvillian matrix in powers of $vt$;
\begin{widetext}
\begin{equation}\label{mk}
\mathbf{M}(t)=(vt)^0\left[
\begin{array}{ccc}
0 & 0 & 0\\
0 & -2\kappa g^2 & -2g\\
0 & 2g & -2\kappa g^2
\end{array}\right]+(vt)^1\left[
\begin{array}{ccc}
0 & -2 & 2\kappa g\\
2 & 0 & 0\\
2\kappa g & 0 & 0
\end{array}\right]+(vt)^2\left[
\begin{array}{ccc}
-2\kappa & 0 & 0\\
0 & -2\kappa & 0\\
0 & 0 & 0
\end{array}\right].
\end{equation} 
\end{widetext}
We note: ($i$) the last term, quadratic in $vt$, generates relaxation of $R_x$ and $R_y$ but does not directly affect $R_z$, ($ii$) for $g=0$ any state on the symmetry axis between the north/south poles is a steady state, ($iii$) for $\kappa=0$ the time-dependence is linear and the second term contains the elements $\pm2vt$ which result in identifying the slope $\alpha=\pi$~\cite{damski2006adiabatic}, and ($iv$) the $\kappa$-dependent elements of the second term clearly describe non-unitary evolution as they appear with the same sign. The fact that $\mathbf M(t)$ contains both a linear and quadratic time-dependence could hint that the rate-of-change should not be taken linear as for the closed case, or in the visual example of Fig.~\ref{fig1}. However, the quadratic time-dependence only enters on the diagonals and does not generate direct couplings between the Bloch vector components. For the linear term we expect that $\alpha=\alpha(\kappa g)$, and  furthermore that $\lim_{\kappa\rightarrow0}\alpha(\kappa g)=\pi$. We also know that we must have $\alpha(\kappa g)<\pi$ for non-zero $\kappa$. 

\begin{figure}
\includegraphics[width=8cm]{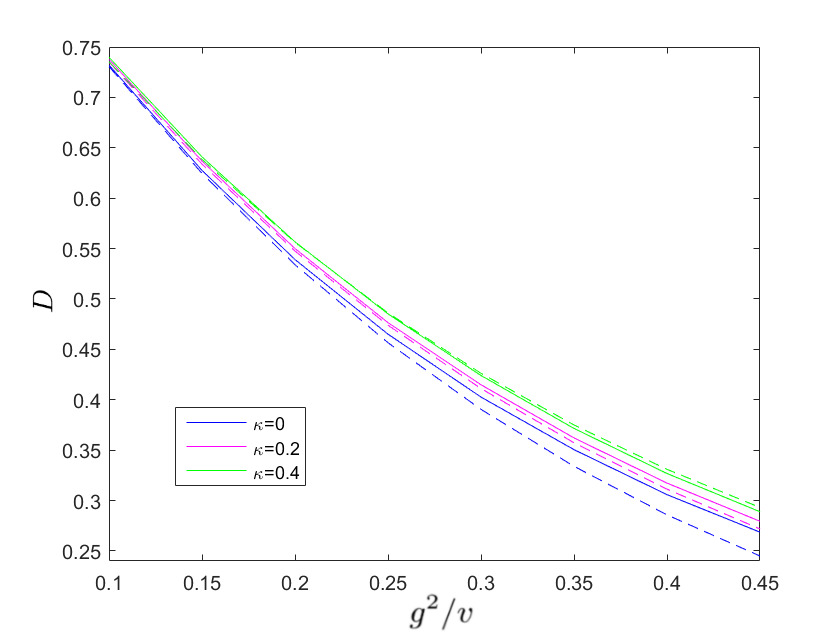} 
\caption{(Color online) The amount of excitations, measured by the trace distance(\ref{trdist2}), as a function of $g^2/v$ for three different loss rates $\kappa$. Solid lines give the results according to the AI-approximation with the rate-of-change(\ref{alphaeq}), while the dashed ones are the corresponding numerical results obtained from direct numerical integration of~(\ref{lzlind}). Note that the agreement gets better the larger $\kappa$ is. The closed case represents the situation studied in Ref.~\cite{damski2005simplest,damski2006adiabatic}. The interval of integration $[t_i,t_f]$ is the same as in Fig.~\ref{fig4}.}
\label{fig6}
\end{figure}

Since the open LZ problem studied here is not analytically solvable (to the best of our knowledge), we cannot use the same arguing as in Ref.~\cite{damski2006adiabatic} to determine $\alpha(\kappa g)$. Assuming a linear $\kappa g$ dependence, a  least square fit gives
\begin{equation}\label{alphaeq}
\alpha=\pi\left(1-\frac{\kappa g}{2}\right).
\end{equation}
Here it is understood that this formula is valid for $\kappa g\ll1$. The freeze-out time is obtained from
\begin{equation}
\pi\left(1-\frac{\kappa g}{2}\right)\hat t=\frac{1}{2\sqrt{\kappa^2\varepsilon^4(\hat t)+\varepsilon^2(\hat t)}}.
\end{equation}
The analytic expression for $\hat t$ is long and not very informative. We note, however, that $\hat t$ is an increasing function of $\kappa$, which explains the larger amount of excitations generated through the quench in the open compared to the closed LZ problem. The applicability of the AI-approximation, using the parameter~(\ref{alphaeq}), is demonstrated in Fig.~\ref{fig6}. The regime of study is for relatively fast quenches where the KZM is believed to reproduce quantitatively correct predictions. For these parameters, the agreement is very convincing, even for as large loss rates as $\kappa=0.4$. Indeed, for the parameters of the figure the agreement is improved with larger $\kappa$. For even larger $\kappa$, beyond $\sim3$, the behavior is qualitatively different as the evolution is then described by a Zeno effect, see discussion in the concluding remarks in Sec.~\ref{sec5}. 
 
\subsection{Dephasing transverse Ising model}\label{ssec4B}
The open LZ model of the previous subsection is a most simple example demonstrating the ideas of the KZM for open quantum systems, but it is not, in a true sense, describing a proper quantum phase transition. In this subsection we turn to a model, the transverse Ising model~\cite{sachdev2007quantum,suzuki2012quantum}, that indeed hosts a quantum phase transition, both in the open and closed version of the model. Thanks to a set of clever transformations proposed by Dziarmaga in Ref.~\cite{dziarmaga2005dynamics}, the transverse Ising model can be mapped to a set of decoupled LZ problems. Each LZ system describes the dynamics of a single momentum mode, and for the of limit long wavelengths ($k\rightarrow0$) the effective LZ velocity $v$ diverges marking the presence of a critical point and the unavoidable breakdown of adiabaticity. 

\begin{figure}
\includegraphics[width=8cm]{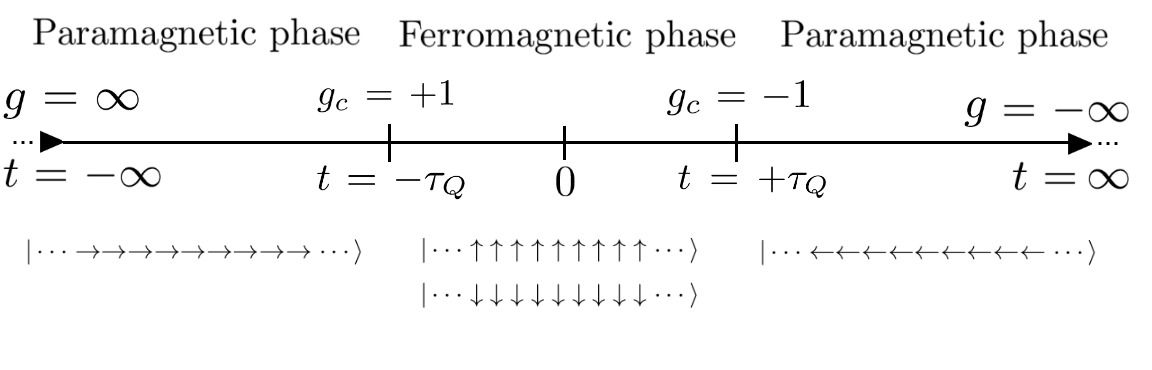} 
\caption{Schematic picture showing the quench order of the transverse Ising model. At $t=-\infty$ we have $g=\infty$ and the system is deep in one of the paramagnetic phases. The ground state is unique and the spectrum gapped. For $t=-\tau_Q$ the system passes the first critical point and enters the symmetry broken ferromagnetic phase. In this phase the $\mathbb{Z}_2$ parity symmetry is broken and the ground state is doubly degenerate. Local excitations consists in domain walls between the two ferromagnetic ground states. At $t=+\tau_Q$ the system goes through the second critical point and ends up in the other paramagnetic phase. }
\label{fig7}
\end{figure}

The transverse Ising model in one dimension for $N$ sites is
\begin{equation}\label{ising}
\hat H_\mathrm{I}=-J\sum_{i=1}^N\left(\hat\sigma_i^z\hat\sigma_{i+1}^z+g\hat\sigma_i^x\right),
\end{equation}
where $J$ is the typical energy scale, $g$ the transverse field strength, and $\hat\sigma_i^\alpha$ ($\alpha=x,\,y,\,z$) the Pauli matrices at site $i$. Throughout we use periodic boundary conditions, $\hat\sigma_1^\alpha=\hat\sigma_{N+1}^\alpha$. For $g=0$ the Hamiltonian is diagonal in the computational basis with the doubly degenerate ferromagnetic ground states $|\uparrow\uparrow\dots\uparrow\rangle$ and $|\downarrow\downarrow\dots\downarrow\rangle$. In the opposite limit $g\rightarrow\pm\infty$, the ground state is paramagnetic (or polarized) $|\!\!\leftarrow\leftarrow\dots\leftarrow\rangle$ or $|\!\!\rightarrow\rightarrow\dots\rightarrow\rangle$. Here $|\!\uparrow\rangle$ and $|\!\downarrow\rangle$ are the eigenstates of $\hat\sigma_z$ with eigenvalues $\pm1$, while $|\!\!\leftarrow\rangle$ and $|\!\!\rightarrow\rangle$ are the eigenstates of $\hat\sigma_x$. At $g_c=\pm1$ there is a continuous QPT between the para- and ferromagnetic phases. Thus, a quench dictated by the time-dependent field strength
\begin{equation}\label{qising}
g(t)=-\frac{t}{\tau_Q}
\end{equation}
drives the system from a paramagnetic phase through a critical point at $g_c=+1$ (i.e. $t=-\tau_Q$) into the symmetry broken ferromagnetic phase and through a second critical point at $g_c=-1$ (i.e. $t=\tau_Q$) back into a symmetric paramagnetic phase. The quench scheme is depicted in Fig.~\ref{fig7}. 

\subsubsection{Mapping to a LZ problem}
Before entering the problem of how dephasing affects the evolution we first discuss the closed system. In Appendix~\ref{appB} we present the details of the mapping of the quenched Ising model to a LZ problem. Here we only record the main results of the mapping. The simplest approach in solving the transverse Ising model is by employing the Jordan-Wigner transformation that casts the Hamiltonian into a problem of free fermions~\cite{sachdev2007quantum}. In the momentum representation the Hamiltonian reads
\begin{equation}
\hat H_\mathrm{I}=2\sum_{k>0}\left[\alpha_k\left(\hat c_k^\dagger\hat c_k-\hat c_{-k}\hat c_{-k}^\dagger\right)+\beta_k\left(\hat c_{-k}^\dagger\hat c_k^\dagger+\hat c_{-k}\hat c_k\right)\right],
\end{equation}
where $\alpha_k=g(t)-\cos(k)$, $\beta_k=\sin(k)$, $\hat c_k^\dagger$ and $\hat c_k$ are the creation and annihilation operators of a fermion with momentum $k$, and the sum runs over only positive momentum modes. As a quadratic Hamiltonian it can be diagonalized by a Bogoliubov transformation which introduces the new fermion operators according to
\begin{equation}\label{bt}
\hat c_k=u_k(t)\hat\gamma_k+v_{-k}^*(t)\hat\gamma_{-k}^\dagger
\end{equation}
and equivalently for $\hat c_k^\dagger$, and to warrant the correct fermionic commutation relations $|u_k(t)|^2+|v_k(t)|^2=1$. The Heisenberg equations give two coupled equations for the Bogoliubov amplitudes as
\begin{equation}\label{BdG}
i\frac{d}{dt}\!\left[\!\begin{array}{c}
u_k(t)\\ v_k(t)\end{array}\!\right]=2\!\left[\begin{array}{cc}
g(t)-\cos(k) & \sin(k)\\ \sin(k) & -g(t)+\cos(k)\end{array}\!\right]\!\!\left[\!\begin{array}{c}
u_k(t)\\ v_k(t)\end{array}\!\right]\!,
\end{equation}
which defines a $2\times2$ $k$- and  time-dependent Hamiltonian $\hat h_\mathrm{LZ}^{(k)}(t)$. With the substitution~\cite{dziarmaga2005dynamics}
\begin{equation}
\tau=\frac{2\tau_Q\sin(k)}{g_0}\left[\frac{t}{\tau_Q}+\cos(k)\right],\hspace{0.5cm}v=\frac{g_0^2}{2\tau_Q\sin^2(k)}
\end{equation}
the equations of motions for the amplitudes become
\begin{equation}
i\frac{d}{d\tau}\!\left[\!\begin{array}{c}
u_k(\tau)\\ v_k(\tau)\end{array}\!\right]=2\!\left[\begin{array}{cc}
-v\tau & g_0\\ g_0 & v\tau\end{array}\!\right]\!\left[\!\begin{array}{c}
u_k(\tau)\\ v_k(\tau)\end{array}\!\right]\!,
\end{equation}
i.e. they attain the familiar form of a LZ problem with a corresponding Hamiltonian $\hat H_\mathrm{LZ}^{(k)}(\tau)$. Thus, we have reduced the analysis of the full time-dependent transverse Ising model to a problem of solving a set of LZ models, one for each momentum mode $k$. 

\subsubsection{Universal scaling of excitations}
For the regular LZ sweep from $\tau=-\infty$ to $\tau=+\infty$, as already mentioned, the system crosses two critical points. We thereby consider the time interval $t\in(-\infty,0]$ to start with, and briefly discuss the second scenario later. The final time $t_f=0$ is assumed to be within the adiabatic regime such that the LZ formula~(\ref{lzformula}) should be applicable. One point to notice is that the LZ crossing instant $\tau=0$ does not coincide with the actual crossing of the critical point at $t=\tau_Q$; it is shifted by $-\tau_Q\cos(k)$. Furthermore, note that for the low energy modes, corresponding to small $k$, the rate $v$ diverges, marking the breakdown of adiabaticity in the vicinity of the critical points.  It is mainly these long wave-length modes being excited during the quench, and we may expand $\sin^2(k)\approx k^2$. By identifying $g_0^2/v=2\tau_Qk^2$ we estimate the excitation from the LZ formula~(\ref{lzformula}),
\begin{equation}
P_\mathrm{LZ}^{(k)}=e^{-2\pi\tau_Qk^2}.
\end{equation}
The total amount of excitations is then evaluated as
\begin{equation}\label{isingdef}
n_D=\sum_{k}P_\mathrm{LZ}^{(k)}\approx\frac{1}{2\pi}\frac{1}{\sqrt{2\tau_Q}}.
\end{equation}
Going back to the KZ prediction~(\ref{defdens}), and using the transverse Ising exponents $z=\nu=1$, we see that it exactly reproduce the above result~(\ref{isingdef})~\cite{dziarmaga2005dynamics}.

\begin{figure}
\includegraphics[width=8cm]{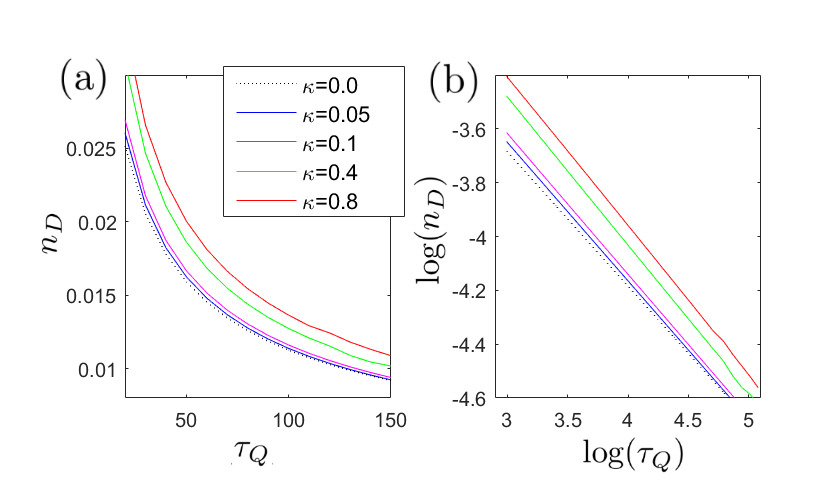} 
\caption{(Color online) The density of defects $n_D$ (a) accumulated through the quench from $t=-\infty$ to $t=0$, and for different decay rates $\kappa$ (see inset). Clearly, the larger $\kappa$ the more excitations. The right panel (b) displays the same but as a log-log plot. The defect density scales with an exponent $\mu$ in all cases (see Tab.~\ref{tab1}), which interestingly is modified by the dephasing. For the numerics, the initial time $t_i=4\tau_Q$ which is deep in the adiabatic regime, and the number of sites $N=512$.}
\label{fig8}
\end{figure}

With the energy dephasing jump operator $\hat L=\hat h_\mathrm{LZ}(t)$, the transformation to fermions in the momentum representation is still straightforward. The total state $\hat\rho(\tau)=\bigotimes_k\hat\rho^{(k)}(\tau)$, where, for each $k$,
\begin{widetext}
\begin{equation}\label{lzlindk}
\partial_t\hat\rho^{(k)}(t)=i\left[\hat\rho^{(k)}(t),\hat h_\mathrm{LZ}^{(k)}(t)\right]
\displaystyle{+\kappa\left(2\hat h_\mathrm{LZ}^{(k)}(t)\hat\rho^{(k)}(t)\hat h_\mathrm{LZ}^{(k)}(t)-\hat h_\mathrm{LZ}^{(k)2}(t)\hat\rho^{(k)}(t)-\hat\rho^{(k)}(t)\hat h_\mathrm{LZ}^{(k)2}(t)\right).}
\end{equation}
\end{widetext}
Thus, for each momentum mode the amount of excitations is derived as in the previous subsection, and their sum gives the total excitations $n_D$. The results of a numerical calculation are presented in Fig.~\ref{fig8} showing $n_D$ for different $\kappa$'s and as a function of $\tau_Q$. The amount of excitations increases with $\kappa$ as anticipated. Universality implies a power-law scaling, which is also found as is evident from the log-log plot, Fig.~\ref{fig8} (b). The interesting result is that the exponent is altered by the dephasing. For mean-field models~\cite{nagy2011critical,baumann2011exploring} and quantum critical models~\cite{patane2008adiabatic,patane2009adiabatic} it has been demonstrated that critical exponents may change due to coupling to an environment. The numerically extracted exponents, $n_D\propto\tau_Q^\mu$, for the examples of Fig.~\ref{fig8} are listed in Tab.~\ref{tab1}.

\begin{table}
\begin{center}
\begin{tabular}{cc}
\Xhline{3\arrayrulewidth}
\hspace{1cm}$\kappa$\hspace{1cm} & \hspace{1cm}$\mu$\hspace{1cm} \\
\hline
0 & -0.5 \\
0.05 & -0.504 \\
0.1 & -0.523 \\
0.4 & -0.541 \\
0.8 & -0.544 \\
\Xhline{3\arrayrulewidth}
\end{tabular}
\caption{Numerically estimated KZ exponents $\mu$, $n_D\propto\tau_Q^\mu$ for different $\kappa$. The $\mu$'s have been calculated from a least square fit from the data of Fig.~\ref{fig8} (b), and the numerically obtained error is around 1$\%$ for all examples. Convergence of the results have been checked, both in respect to the integration interval and the system size. }\label{tab1}
\end{center}
\end{table}

\subsubsection{Correlation function}
For a quench ending at $g(t_f)=0$ in the ferromagnetic phase we expect excitations in terms of domain walls (also called kinks in one dimension). According to the result above, the KZM predicts a correlation length $\xi\propto\sqrt{\tau_Q}$ for the closed Ising model. For non-zero $\kappa$ we still expect domain wall excitations as the Lindblad jump operator commutes with the Hamiltonian for every time instant. Nevertheless, we saw that the exponents are modified by the dephasing and should accordingly affect the correlation length.

The characteristic length $\xi$ between the domain wall excitations is extracted from the correlator
\begin{equation}\label{corrf0}
C_R^{z}\equiv\langle\hat\sigma_i^z\hat\sigma_{i+R}^z\rangle-\langle\hat\sigma_i^z\rangle\langle\hat\sigma_{i+R}^z\rangle=\langle\hat\sigma_i^z\hat\sigma_{i+R}^z\rangle,
\end{equation}
where we have used that $\langle\hat\sigma_i^z\rangle=0$ from symmetry. The derivation of the correlation function is reproduced in Appendix~\ref{appC} following Ref.~\cite{dziarmaga2010dynamics2}. The crucial observation is that the correlation function can be written as a determinant of a matrix of pair correlators;
\begin{equation}\label{corrf}
|C_R^z|=\sqrt{|\mathrm{det}(Q_R)|},
\end{equation}
where
\begin{equation}
Q_R=\left[\begin{array}{cccc}
G_{11} & G_{12} & \hdots & G_{1R}\\
G_{21} & G_{22} & \hdots & G_{2R}\\
\vdots & \vdots & \ddots & \vdots\\
G_{R1} & G_{R2} & \hdots & G_{RR}
\end{array}\right]
\end{equation}
and
\begin{equation}\label{gmatrix}
G_{ij}=\left[\begin{array}{cc}
\langle\hat A_{i+1}\hat A_{j+1}\rangle & \langle\hat B_{i}\hat A_{j+1}\rangle\\
\langle\hat A_{i+1}\hat B_{j}\rangle & \langle\hat B_{i}\hat B_{j}\rangle
\end{array}\right],
\end{equation}
with $\hat A_i=\left(\hat c_i^\dagger+\hat c_i\right)$ and $\hat B_i=\left(\hat c_i^\dagger-\hat c_i\right)$. The correlators of Eq.~(\ref{gmatrix}) can be expressed in terms of the Bloch vector components (see Apprendix~\ref{appC}), for example
\begin{equation}\label{corr2}
\langle\hat B_i\hat A_j\rangle=-\frac{1}{N}\sum_k\left[R_k^z\cos(k(i-j))+R_k^x\sin(k(i-j))\right].
\end{equation}
As time progresses, the different Bloch vectors will depart one another causing an intrinsic dephasing and a decay of the correlations. This dephasing is manifested over a length scale $L=\sqrt{\tau_Q}\log\tau_Q$~\cite{dziarmaga2010dynamics2}. However, since the dephasing occurs after some time $T_2$ it will not be seen immediately, for example at the critical point it is not established but it is at $g=0$.

\begin{figure}
\includegraphics[width=8cm]{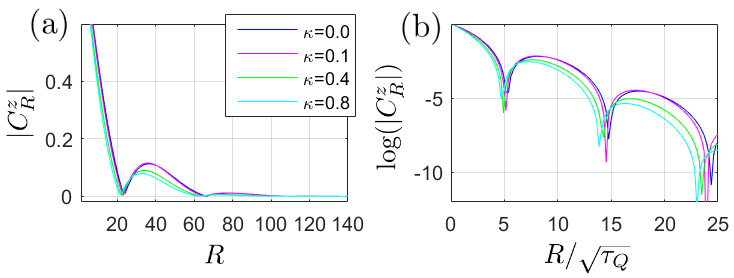} 
\caption{(Color online) The absolute value of the $z$-correlation function (a) and its logarithm (b) as a function of the distance $R$. The curves give the results for different $\kappa$ values, and the final time is such that $g(t_f)=0$. The quench time $\tau_Q=20$, and $t_i=4\tau_Q$. The distances between local minima, especially visible in (b), give the typical length $\xi$ that for $\kappa=0$ should scale as $\sqrt{\tau_Q}$ according to the KZM. This has also been confirmed here but is not explicitly shown as we only consider one time $\tau_Q$. The effect of the dephasing, i.e. non-zero $\kappa$, is not very dominant; a slight shorter correlation length $\xi$ is found.}
\label{fig9}
\end{figure}

The numerically extracted correlation function for different rates $\kappa$ is shown in Fig.~\ref{fig9}. In the ferromagnetic phase, in the presence of a domain wall $C_R^z$ should flip sign, but since the analytical expression~(\ref{corrf}) only gives the absolute vale we cannot see such a sign change. However, it is clear that $|C_R^z|$ becomes very small for periodic $R$ values. The length between the minima gives $\xi$. The scaling $\xi\propto\sqrt{\tau_Q}$ for $\kappa=0$ has been confirmed numerically by calculating the correlation function for different $\tau_Q$'s. According to Tab.~\ref{tab1}, the environmental dephasing slightly changes the KZ exponent $\mu$ which is also seen for the correlation function, particularly evident in the log plot (b).

We have also analyzed the quench when extended to $t=+\infty$ such that the system is driven through two critical points, one at $g=-1$ as above and one at $g=+1$. Naturally, at both critical points the spectrum is gapless and we may therefore envision the scheme as a sort of interferometer; in the vicinity of $g=-1$ different energy eigenstates get populated, and later at $g=+1$ when the gap closes again the different states mix and the interference loop(s) is closed. In the setting of two LZ crossings, such an interferometer is called a Landau-Zener-St\"uckelberg interferometer~\cite{shevchenko2010landau}. The external dephasing appearing whenever $\kappa\neq0$ could well affect the interferences and thereby the resulting amount of excitations. This dephasing, stemming from incoherent time evolution, is of course different from the coherent intrinsic one deriving from different momentum modes $k$ as discussed above. Nevertheless, for the correlation function~(\ref{corrf}), constructed from terms like~(\ref{corr2}), it is found that the external dephasing does not qualitatively alter the behavior of the correlation function, see Appendix~\ref{appC}. In short, the sum in~(\ref{corr2}) can be small either due to canceling terms or if the different momentum $k$ Bloch vectors shrink.


\section{Summary and future directions}\label{sec5}
In their seminal work~\cite{zurek2005dynamics}, Zurek {\it et al.} suggested how the resulting dynamics when systems are driven across quantum critical points can be understood in terms of the AI-approximation. Since then, this KZM has been verified both numerically and experimentally in a variety of systems, but also its limitations have been explored, see for example~\cite{su2013kibble}. In this work we have generalized the KZM as applied to closed quantum systems to open quantum systems, i.e. systems showing driven-dissipative critical behavior. This new type of non-equilibrium quantum PT's has gained much attention lately due to recent experimental progress in the AMO community of ultracold physics. As the field of NESS criticality is still very young not much is known even if the field is developing rapidly. Our results add to the understanding of these systems. Many of the previous results point in the direction that the physics behind NESS PT's bare much in common with equilibrium QPT's, as for example universality, but as we pointed out there exist also differences. In this respect, it is not {\it a priori} clear that teh KZM presented in Ref.~\cite{zurek2005dynamics} can be generalized to NESS PT's. 

The KZM, hinging on the AI-approximation, relies on some fundamental concepts, namely adiabaticity and universal scaling. Thus, for any generalization of the KZM to NESS PT's one must first explore these concepts in terms of open quantum systems. Adiabaticity for open quantum systems, as discussed in Sec.~\ref{adsubsec}, is by now rather well understood, partly thanks to works in adiabatic quantum computing~\cite{sarandy2005adiabatic2,sarandy2005adiabatic,joye2007general,avron2012adiabatic,venuti2016adiabaticity}. The fact that the time-evolution is non-unitary in general (manifested for example as complex Liouvillian eigenvalues), even under extremely slow parameter changes an adiabatic following may imply some change in the system's state. In this work we are interested in the instantaneous steady states which by definition has a zero eigenvalue and adiabatic evolution implies no such relaxation. Even if adiabaticity for open quantum systems is understood to a great extent, the same cannot be said about critical scaling, or in particular the scaling of the eigenvalues in the vicinity of a critical point. The question seems to go back to~\cite{kessler2012dissipative} where it is pointed out that the Liouvillian gap~(\ref{lgap}) must close at the critical point. Numerics is substantially much more difficult when diagonalizing Lindblad equations than simple Hamiltonians, which make finite size scaling explorations of the Liouvillian spectrum harder~\cite{vicentini2017critical}. Nevertheless, there are evidences supporting critical slowing down for open quantum systems, which motivates a KZM approach in order to explore quenches across NESS critical points. The KZM for open quantum systems was discussed in Sec.~\ref{ssec3d}, and we especially pointed out that the relevant time-scales, reaction time and the inverse transition rate (see Fig.~\ref{fig1}), may well depend on the decay rate $\kappa$, which in return implies that also the extent of the impulse regime will depend on $\kappa$.  

Related to the question about adiabaticity for open quantum systems is that of characterizing the amount of non-adiabatic excitations, which was the topic of Sec.~\ref{ssec3C}. As we have emphasized, for driven-dissipative critical quantum systems the steady state $\hat\rho_\mathrm{ss}$ replaces the role played by the ground state in critical closed quantum systems. The KZ scaling in closed systems gives a prediction of the amount of excitations relative the ground state. Since the steady state may be very distinct from the ground state of the Hamiltonian of Eq.~(\ref{lindblad}) we should not define non-adiabatic excitations with respect to that ground state, but instead in relation to the adiabatic instantaneous steady state $\hat\rho_\mathrm{ss}^\mathrm{(ad)}(t)$. For an adiabatic evolution, according to its definition, the time-evolved state will identify $\hat\rho_\mathrm{ss}^\mathrm{(ad)}(t)$ at every instant, and any deviations from it are ascribed non-adiabatic excitations. As argued in Sec.~\ref{ssec3C}, the natural measure is then the trace distance of density operators, Eqs.~(\ref{tr1}) and~(\ref{trdist}). In particular, we showed that for energy dephasing (where the energy eigenbasis can be seen as the most relevant one) the trace distance gives the amount of energetic excitations above the ground state. A new ingredient in the dynamics of open QPT's is the fact that the state typically becomes mixed as time progresses, see Fig.~\ref{fig4}. This typically increases the amount of excitations, which is in agreement with what has been found in earlier studies~\cite{fubini2007robustness,cincio2009dynamics,dutta2016anti}. Thus, this observation provides additional insight into those processes. We did not analyze the regime of very large $\kappa$ in this work. We have, however, explored this limit numerically and we found that the evolution can enter a `quantum Zeno' regime which is conceptually different from the results presented in this work. In the field of quantum coherent control, exploiting the Zeno effect has been analyzed in the past~\cite{maniscalco2008protecting,scala2010stimulated,mathisen2016view}. A more thorough study of this phenomenon by extending it to NESS critical models is left for the future.

With the open LZ problem, analyzed Sec.~\ref{ssec4A}, we verified the applicability of the KZM to such an open quantum systems. When taking into account that the rate-of-change parameter $\alpha$ must be dressed with a $\kappa$-dependence we found very good agreement between the results predicted from the KZM and those obtained from direct numerical integration, see Fig.~\ref{fig6}. The explicit $\kappa$-dependence was not rigorously proven, but motivated both numerically and from arguments based on the form of the Liouvillian matrix $\mathbf{M}(\kappa)$ of Eq.~(\ref{mk}). The KZ scaling for a model supporting a true critical point was demonstrated in Sec.~\ref{ssec4B} for the transverse Ising model exposed to energy dephasing. Critical exponents were extracted numerically and we indeed found a slight $\kappa$-dependence which implies that the openness of the problem may alter the Ising universality class. The shift in the exponents resulted in a slight change also in the density of defects as exhibited in Fig.~\ref{fig9}.

There are several open questions to be addressed, and we see many possible future directions. On a more general level, establishing the scaling of the Liouvillian spectrum in the vicinity of the critical point is certainly an important issue. For example, it was recently shown that a continuous PT is possible for an open quantum system where the steady state is unique, i.e. there is no symmetry breaking accompanying the PT~\cite{hannukainen2017dissipation}. A related question, relevant also for the present work, is the types of excitations in NESS critical systems. For closed systems we know that the symmetries and dimensions determine the character of excitations, e.g. vortices, domain walls, and waves. For Lindblad master equations, on the other hand, symmetries and conserved quantities are not necessarily linked~\cite{albert2014symmetries,albert2016geometry}. In the present work the two studied examples belong to Class I, i.e. representing energy dephasing, where the type of the excitations are not expected to change in comparison to the closed case.  It would therefor definitely be interesting to analyze the other classes as well. Especially since then there need not be any clear link between the steady state and the Hamiltonian eigenestates. A difficulty here is that it seems that models in these classes become substantially more complicated, unless they are in some sense trivial. For our example of the Ising model we found that the excitations were domain walls (kinks) as for the closed system. And we also saw that the correlation function behaved much the same as for the closed Ising model. This suggests that it would be interesting to analyze other type of correlators that are connected to quantum entanglement and not classical correlations as studied here. In such situations we may expect more distinct differences between the open and closed cases. As a final remark, the classification scheme in Sec.~\ref{ssec2B} is limited to single loss channels, and it is unclear whether it makes sense to classify the general case or if there are simply too many qualitatively different scenarios or classes.


\begin{acknowledgements}
We thank Irina Dumitru and Thomas Kvorning for helpful discussions. The Knut and Alice Wallenberg foundation (KAW) and the Swedish research council (VR) are acknowledged for financial support.  
\end{acknowledgements}

\appendix
\section{Liouvillian matrix for general qubit master equation}\label{appA}
In this Appendix we give the most general expression of the Bloch equations for a two-level system. In a proper frame we may take a general two-level Hamiltonian as
\begin{equation}
\hat H_\mathrm{qu}=\left[
\begin{array}{cc}
\varepsilon & g\\
g & -\varepsilon
\end{array}\right].
\end{equation}
Furthermore, we can always decompose the Bloch equations as
\begin{equation}
\partial_t\mathbf{R}=\mathbf{M}_H\mathbf{R}+\sum_i\left(\mathbf{M}_{L_i}\mathbf{R}+\mathbf{b}_i\right).
\end{equation}
Here,
\begin{equation}
\mathbf{M}_H=2\left[
\begin{array}{ccc}
0 & -\varepsilon & 0\\
\varepsilon & 0 & -g\\
0 & g & 0
\end{array}\right]
\end{equation}
is the part of the Liouvilian matrix deriving from the Hamiltonian part, 
\begin{widetext}
\begin{equation}
\mathbf{M}_{L_i}=\kappa_i\left[
\begin{array}{ccc}
-\left(|L_{yi}|^2+|L_{zi}|^2\right) & L_{xi}L_{yi}^*+iL_{1i}L_{zi}^* & L_{zi}^*L_{xi}-iL_{1i}L_{yi}^*\\
L_{xi}^*L_{yi}-iL_{1i}L_{zi}^* & -\left(|L_{xi}|^2+|L_{zi}|^2\right) & L_{yi}L_{zi}^*+iL_{1i}L_{xi}^*\\
L_{zi}L_{xi}^*-iL_{1i}L_{yi}^* & L_{yi}^*L_{zi}-iL_{1i}L_{xi}^* & -\left(|L_{xi}|^2+|L_{yi}|^2\right)
\end{array}\right]+c.c.
\end{equation} 
\end{widetext}
is the contribution to $\mathbf M$ resulting from the jump operator $\hat L_i$, and the corresponding pump term
\begin{equation}
\mathbf b_i=2i\kappa_i\left[
\begin{array}{c}
L_{yi}^*L_{zi}\\
L_{zi}^*L_{xi}\\
L_{xi}^*L_{yi}
\end{array}\right]+c.c.\,.
\end{equation} 
The jump operators are on the general form $\hat L_i=L_{1i}\mathbb{I}+L_{xi}\hat\sigma_x+L_{yi}\hat\sigma_y+L_{zi}\hat\sigma_z$. We directly note that for hermitian jump operators $\hat L_i$, the corresponding pump term vanishes, i.e. $\mathbf b_i=0$. The pump term also disappear when only a single $L_{\alpha i}$ ($\alpha=x,\,y,\,z$) is non-zero, i.e. dephasing in a given basis.

\section{Quenched transverse Ising model as a set of decoupled LZ problems}\label{appB}
In this Appendix, following Ref.~\cite{dziarmaga2005dynamics}, we present the derivation of how the transverse Ising Hamiltonian~(\ref{ising}), with a linear quench in the field strength $g(t)$, can be mapped onto a set of decoupled LZ problems. 

The method of solving the Ising model can be found in many text books, see for example~\cite{sachdev2007quantum,suzuki2012quantum}. The standard approach is to first apply the Jordan-Wigner transformation defined as
\begin{equation}\label{jw}
\begin{array}{l}
\hat\sigma_i^x=1-2\hat c_i^\dagger\hat c_i,\\ \\
\hat\sigma_i^z=-\left(\hat c_i^\dagger+\hat c_i\right)\prod_{j<i}\left(1-2\hat c_i^\dagger\hat c_i\right),
\end{array}
\end{equation}
where the $\hat c_i$'s are regular fermionic operators. Thus, with Eq.~(\ref{jw}) the Ising Hamiltonian is transformed into one of spinless fermions. After some algebra, especially handling the fermionic ``string'' arising from the product of fermion operators in~(\ref{jw}), one finds
\begin{equation}
\begin{array}{lll}
\hat H_\mathrm{I} & = & \displaystyle{J\sum_i\left[\hat c_i^\dagger\hat c_{i+1}+\hat c_{i+1}^\dagger\hat c_i+\hat c_i^\dagger\hat c_{i+1}^\dagger+\hat c_{i+1}\hat c_i\right.}\\ \\
& & \left.-g\left(2\hat c_i^\dagger\hat c_i+1\right)\right].
\end{array}
\end{equation}
As a quadratic Hamiltonian it describes free fermions on a one dimensional lattice. The pairing terms break, however, particle conservation. After going to the momentum representation, and rewrite the sum only over positive momentum terms, the fermionic Hamiltonian becomes
\begin{equation}
\begin{array}{lll}
\hat H_\mathrm{I} & = & \displaystyle{2J\sum_{k>0}\left[(g-\cos(k))\left(\hat c_k^\dagger\hat c_k-\hat c_{-k}\hat c_{-k}^\dagger\right)\right.}\\ \\
& & \left.\sin(k)\left(\hat c_k^\dagger\hat c_{-k}^\dagger+\hat c_{-k}\hat c_k\right)\right].
\end{array}
\end{equation}
This Hamiltonian is diagonalized by introducing the Boloiubov modes $\hat\gamma_k$ according to
\begin{equation}
\hat c_k=u_k\hat\gamma_k+v_{-k}^*\hat\gamma_{-k}^\dagger,
\end{equation}
where $|u_k|^2+|v_k|^2=1$ to guarantee that the new modes are fermionic. With the present model one further has the relations $u_k=u_{-k}$ and $v_k=-v_{-k}$. The coefficients that diagonalize the Hamiltonian are given by the Bogoliubov-deGennes' eigenvalue equation
\begin{equation}
\epsilon_\pm\left[
\begin{array}{c}
u_k\\ v_k
\end{array}\right]=2
\left[\begin{array}{cc}
g-\cos(k) & \sin(k)\\
\sin(k) & -(g-\cos(k))
\end{array}\right]\left[
\begin{array}{c}
u_k\\ v_k
\end{array}\right],
\end{equation}
where the eigenvalues are $\epsilon_\pm=\pm2\sqrt{\left[g-\cos(k)\right]^2+\sin^2(k)}$. The spectrum is now given by $\varepsilon_k=\epsilon_+$, and for $k=0$ the gap closes at $g=1$ and for $k=\pi$ it closes at $g=-1$ representing the two critical points. Using the expression for the dispersion together with finite scaling one finds the exponents $z=\nu=1$.

To generalize the scheme above to a time-dependent problem we assume that the Bogoliubov coefficients are time-dependent, see Eq.~(\ref{bt}). To find an equation of motion for the coefficients we start with the Heisenberg equations for the operators $\hat c_k(t)$, and then do the Bogoliubov transformation to the new fermion operators. Using that the different $k$-mode operators are independent one derives, after some algebra, the dynamical Bogoliubov-deGennes equation~(\ref{BdG}).

\section{Derivation of the correlation Ising function}\label{appC}
Just like the previous Appendix, here we briefly sketch the derivation of the correlation function~(\ref{corrf}) as presented in Ref.~\cite{sachdev2007quantum,suzuki2012quantum}. 

In terms of spinless fermions the product of Pauli $\hat\sigma_z$ matrices can be expressed after some manipulations as
\begin{equation}
\hat\sigma_i^z\hat\sigma_{i+R}^z\!=\!\left(\!\hat c_i^\dagger-\hat c_i\!\right)\!\!\prod_{j=i+1}^{i+R-1}\!\!\left(\!\hat c_j^\dagger+\hat c_j\!\right)\!\left(\!\hat c_j^\dagger-\hat c_j\!\right)\!\left(\!\hat c_{j+R}^\dagger+\hat c_{j+R}\!\right)\!.
\end{equation}
After introducing new operators
\begin{equation}
\hat A_i=\left(\hat c_i^\dagger+\hat c_i\right),\hspace{0.5cm}\hat B_i=\left(\hat c_i^\dagger-\hat c_i\right),
\end{equation}
as in the main text, the correlator can be written as
\begin{equation}
C_R^z=\!\langle\hat\sigma_i^z\hat\sigma_{i+R}^z\rangle\!=\!\langle\hat B_i\hat A_{i+1}\hat B_{i+1}\cdots\hat A_{i+R-1}\hat B_{i+R-1}\hat A_{i+R}\rangle.
\end{equation}
The expectation is evaluated be utilizing Wick's theorem, that is we can rewrite an expectation of products of operators as products of expectations of pairs of operators;
\begin{equation}\label{wick}
\langle\hat O\!\cdots\!\hat O_{2i}\rangle\!=\!\sum_p\!(-1)^p\langle\hat O_{w_1}\hat O_{w_2}\rangle\!\langle\hat O_{w_3}\hat O_{w_4}\rangle\!\cdots\!\langle\hat O_{w_{2i-1}}\hat O_{w_{2i}}\rangle\!.
\end{equation}
Here the sum is over all permutations of the numbers $1,\,2,\,\dots,\,2i$, and the subscripts $w_i$ are labeling the different permutations. The expectation~(\ref{wick}) can be identified as a Pfaffian of a $2i\times2i$ matrix where the subscripts mark the matrix indices. But since the Pfaffian can be expressed as the square root of the determinant of the matrix, the correlation function~(\ref{corrf}) follows. 

\bibliographystyle{apsrev}
\bibliography{./bibli}
\end{document}